\documentclass[aps,pra,twocolumn,a4paper,showpacs,superscriptaddress,floatfix,10pt]{revtex4}
\usepackage[pdftex]{graphicx,color}
\usepackage{dcolumn}
\usepackage{bm}

\usepackage{amsfonts}
\usepackage{amssymb}
\usepackage{amsmath}
\usepackage{amsthm}

\usepackage{subfigure}
\usepackage{rotate}

\usepackage{pdfsync}

\usepackage[pdftex]{hyperref}

\newcommand{\noshow}[1]{}

\let\oldmarginpar\marginpar
\renewcommand\marginpar[1]{\-\oldmarginpar[\raggedleft\tiny #1]%
{\raggedright\tiny #1}}

\DeclareMathOperator{\Tr}{Tr}

\newcommand{\bra}[1]{\langle#1|}
\newcommand{\ket}[1]{|#1\rangle}

\graphicspath{{figs/}}

\begin{document}

\title{Many-body localization beyond eigenstates in all dimensions}

\author{A. Chandran}
\affiliation{Perimeter Institute for Theoretical Physics, Waterloo, Ontario N2L 2Y5, Canada}   \email{achandran@perimeterinstitute.ca}

\author{A. Pal}
\affiliation{Rudolf Peierls Centre for Theoretical Physics, Oxford University, Oxford OX1 3NP, UK}
\email{arijeet.pal@physics.ox.ac.uk}

\author{C. R. Laumann}
\affiliation{Department of Physics, University of Washington, Seattle, WA 98195, USA}
\affiliation{Perimeter Institute for Theoretical Physics, Waterloo, Ontario N2L 2Y5, Canada} 

\author{A. Scardicchio}
\affiliation{Abdus Salam ICTP, Strada Costiera 11, 34151 Trieste, Italy}
\affiliation{INFN, Sezione di Trieste, Via Valerio 2, 34127 Trieste, Italy}
\date{\today}

\begin{abstract}
Isolated quantum systems with quenched randomness exhibit many-body localization (MBL), wherein they do not reach local thermal equilibrium even when highly excited above their ground states.
It is widely believed that individual eigenstates capture this breakdown of thermalization at finite size.
We show that this belief is false in general and that a MBL system can exhibit the eigenstate properties of a thermalizing system.
We propose that localized approximately conserved operators (l$^*$-bits) underlie localization in such systems. 
In dimensions $d>1$, we further argue that the existing MBL phenomenology is unstable to boundary effects and gives way to l$^*$-bits.
Physical consequences of l$^*$-bits include the possibility of an eigenstate phase transition within the MBL phase unrelated to the dynamical transition in $d=1$ and thermal eigenstates at all parameters in $d>1$.
Near-term experiments in ultra-cold atomic systems and numerics can probe the dynamics generated by boundary layers and emergence of l$^*$-bits.
\end{abstract}

\maketitle

\newcommand{\Hbulk}{H_{b}}
\newcommand{\Hbdy}{H_{\partial}}
\newcommand{\Hint}{H_{int}}
\newcommand{\OO}{\mathcal{O}}
\newcommand{\lstar}{\tau^{*z}_{\mathbf{i}}}

\section{Introduction}

The development of synthetic quantum many-body systems has rejuvenated interest in the foundations of statistical mechanics. 
In particular, under what conditions does an isolated system establish local thermal equilibrium? 
While the general conditions are unknown, there is growing evidence that strong quenched disorder can localize interacting quantum particles and thereby prevent the exchange necessary for equilibration \cite{anderson1958absence,basko2006metal,gornyi2005interacting,oganesyan2007localization,Nandkishore:2015aa,Altman:2015aa,Eisert:2015aa}.
The primary observational signature of such `many-body localization' (MBL) is the persistence of local memory of initial conditions: 
this has been established in various lattice models theoretically and numerically \cite{basko2006metal,gornyi2005interacting,oganesyan2007localization,Monthus:2010vn,vosk2013dynamical,Pekker:2014aa,znidaric2008many,pal2010mb,Bardason2012,serbyn2013universal,Swingle:2013aa,Serbyn:2014aa,iyer2013many,kjall2014many,laumann2014many,Li:2015aa,luitz2015many,Tang:2015th,Singh:2016aa} and has also been observed in state-of-the-art experiments in ultracold atomic \cite{Schreiber:2015aa,Kondov:2015aa,Bordia:2016aa,Choi:2016aa} and trapped ion systems \cite{Smith:2015aa}. 
Although the understanding of MBL is in its infancy, it has far-reaching consequences for quantum computation \cite{Serbyn:2014ek,Yao:2015aa,Bahri:2015aa}, unconventional quantum phase transitions \cite{Potter:2015ab,Chandran:2015ac,Vosk:2015aa} and out-of-equilibrium phases of matter \cite{Huse:2013aa,Pekker:2014aa,Chandran:2014aa,Bahri:2015aa,Potter:2015aa,von-Keyserlingk:2016aa,Else:2016aa}. 

As the distinction between thermal and MBL phases of matter lies in their long-time dynamics, many recent studies have focused on the structure of many-body eigenstates, which \emph{prima facie} probe infinite time behavior. 
There are good reasons for this.
Thermalization in classical Hamiltonian systems emerges from the unbiased exploration of equal energy surfaces in phase space \cite{Reif:1965aa}. 
After quantization, the closest analog of the fixed energy surfaces are provided by the discrete collection of many-body eigenstates. 
The eigenstate thermalization hypothesis (ETH) holds that these eigenstates, like the classical stationary states, are as random as possible subject to the global energy constraint \cite{deutsch1991quantum,srednicki1994chaos,Tasaki:1998aa,Rigol:2008bh}. 
In particular, the expectation values of few-body operators within individual eigenstates coincide with the thermal ones.
There is a growing body of theoretical and numerical work supporting this hypothesis \cite{Rigol:2008bh,Khlebnikov:2013aa,Beugeling:2014aa,Sorg:2014aa,Steinigeweg:2014aa,Mondaini:2016aa}.

On the other hand, the seminal perturbative work of Basko, Aleiner and Altshuler suggests that many-body eigenstates in the MBL phase are localized in Fock space \cite{basko2006metal}.
As a vertex in Fock space is labelled by occupation numbers, several groups conjectured that the localized eigenstates are labelled by a complete set of dressed occupation numbers or `l-bits' in addition to the energy \cite{huse2013phenomenology,serbyn2013local}.
This conjecture has been rigorously proven in certain one-dimensional spin chains in the limit of strong disorder \cite{imbrie2014many} and many numerical and perturbative constructions of the l-bits are now available \cite{chandran2015constructing,ros2015integrals,Monthus:2016aa,Rademaker:2016aa}.

The eigenstate perspective has proven extremely useful for studying localization.
In this view, the many-body localization transition appears as an eigenstate phase transition between ETH-satisfying and ETH-violating states.
Such ETH-violating eigenstates have efficient tensor product representations with area law entanglement entropy \cite{pal2010mb,Bauer:2013jw,Pekker:2014ab,Chandran:2015aa,Friesdorf:2015aa,Khemani:2015aa,yu2015finding}.
Further, they can exhibit quantum and symmetry breaking orders disallowed by statistical mechanics \cite{Huse:2013aa,Pekker:2014aa,Chandran:2014aa,Bahri:2015aa,Yao:2015aa}.

In this article, we critique the eigenstate perspective and argue that eigenstates need not detect the breakdown of thermalization.
It is entirely consistent, especially in $d>1$, for the MBL phase to have eigenstates which satisfy ETH at all finite sizes so long as the spectral functions vanish appropriately in the thermodynamic limit.
Fundamentally, the limits $L\to \infty$ and $t\to \infty$ do not commute: 
eigenstates are well-defined in the limit $t \to \infty$ followed by $L\to \infty$, while the dynamical phase of matter is well-defined in the opposite order of limits. 
The incorrect order of limits may lead to the conclusion that MBL does not exist in $d>1$ and misidentify the location of the dynamical transition in $d=1$.

\begin{figure}[tbp]
\begin{center}
\includegraphics[width=0.7\columnwidth]{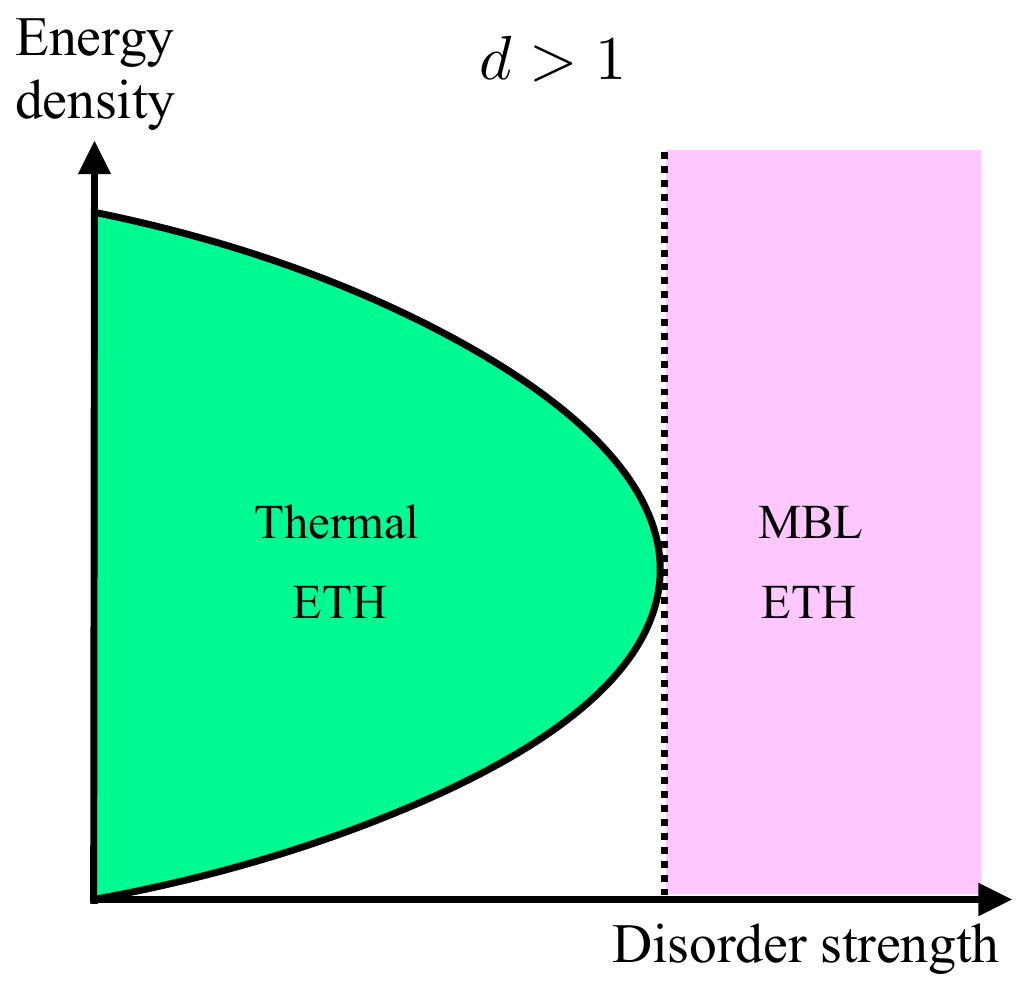}
\caption{Phase diagram in $d>1$ indicating dynamical phases and eigenstate properties. The fully MBL phase is shaded pink and described by l$^*$-bits at all disorder strengths. ETH is satisfied everywhere in the phase diagram with the weight in the spectral function vanishing as $L\to \infty$ in the fully MBL phase. The black line marks the delocalization transition.}
\label{Fig:PhaseDiagramsdgreatone}
\end{center}
\end{figure}

\begin{figure}[tbp]
\begin{center}
\includegraphics[width=0.7\columnwidth]{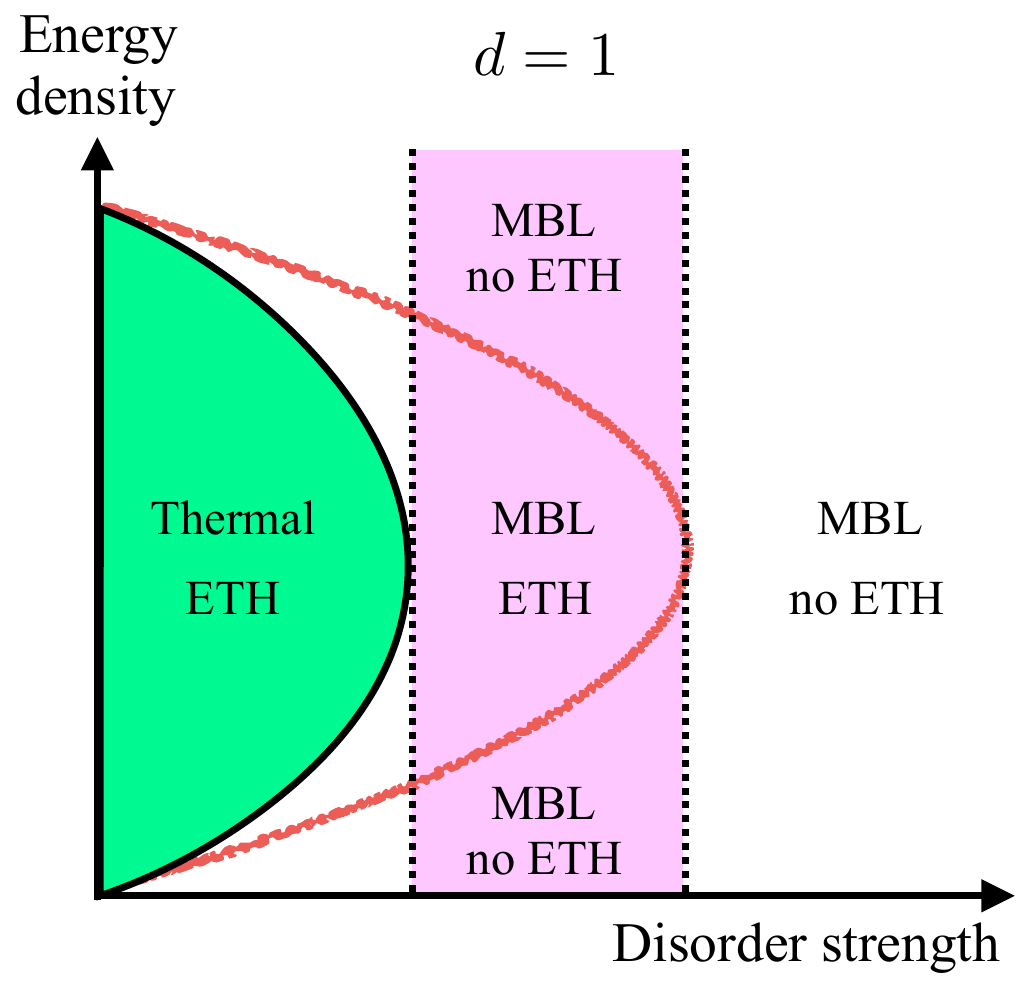}
\caption{Phase diagram in  $d=1$ indicating dynamical phases and eigenstate properties. At strong disorder, the system is fully MBL with l-bits and violates ETH. At intermediate disorder (shaded pink), the system is fully MBL with l$^*$-bits. The red line marks the eigenstate phase transition between ETH and non-ETH states in this regime. The true delocalization transition is denoted by the black line and can lie entirely to the left of the eigenstate phase transition. }
\label{Fig:PhaseDiagrams}
\end{center}
\end{figure}

We propose a refined phenomenology of MBL in terms of approximately conserved quasi-local operators (\mbox{l$^*$-bits}), generalizing the strictly conserved l-bits of previous work.
Such l$^*$-bit systems satisfy ETH at all energy densities in $d>1$ and in an energy density window in $d=1$.
The corresponding phase diagrams are shown in Figs.~\ref{Fig:PhaseDiagramsdgreatone} and~\ref{Fig:PhaseDiagrams}.
In $d>1$, we expect ETH to hold throughout the phase diagram.
In $d=1$, as the MBL phase is described by l-bits at strong disorder, there can be an eigenstate phase transition within the MBL phase where the description changes from l-bits to l$^*$-bits.
In this scenario, the eigenstate phase transition that has been observed in many numerical studies \cite{kjall2014many,luitz2015many,Serbyn:2015aa,Baygan:2015aa} does not coincide with the true dynamical transition.
We suggest numerical tests of possibility.
Like the l-bit ansatz, the l$^*$-bit ansatz only describes localized systems without many-body mobility edges.

In Sec.~\ref{Sec:BoundaryInstab}, we provide evidence that the refined phenomenology is necessary in higher dimensions.
Specifically, in $d>2$, we show that the l-bit structure is always unstable to the inclusion of a thermal boundary layer at finite size, even as the dynamics remain localized in the thermodynamic limit.
The combined system is described by l$^*$-bits and its eigenstates satisfy ETH.
As the l-bit structure is not robust to boundary perturbations even at strong disorder, it is clearly not a stable characterization of the thermodynamic phase. 
The $d=2$ case is marginal for the specific boundary instability we consider, although we believe that l$^*$-bits are generic here as well. 

Our analysis of the thermal layer coupled to a strongly localized bulk is of independent interest in the study of MBL systems (see for example Refs.~\cite{Nandkishore:2014ys,Johri:2015mz,Huse:2015ys,Hyatt:2016aa}). 
It would be very interesting to measure the dynamical influence of a thermal boundary on a localized bulk. 
These experiments are readily accessible with existing technology in ultracold atomic systems \cite{Schreiber:2015aa,Greif953,Bordia:2016aa,Choi:2016aa}.
We take up such experimental considerations in Sec.~\ref{Sec:Experiments}.

\section{Background}
\label{Sec:Background}

Consider an isolated quantum system with Hamiltonian $H$ prepared in the initial state $\ket{\psi}$. 
The dynamical phase is defined by the behavior of expectation values of local observables $\OO$ at late times in the thermodynamic limit. 
More precisely, the objects of interest are
\begin{align}
\label{Eq:LimitOrder}
	\lim_{t\to\infty} \lim_{L\to\infty} \bra{\psi}\OO(t)\ket{\psi} 
\end{align}
In a thermalizing phase, for a large class of initial states $\ket{\psi}$ \footnote{The initial states should be statistically translationally invariant. Eq.~\eqref{Eq:LimitOrder} does not hold for initial states that contain gradients in conserved densities on the scale of the system.}, such expectation values agree with those in the appropriate Gibbs ensemble,
\begin{align}
	\label{eq:thermaldefn}
	\lim_{t\to\infty} \lim_{L\to\infty} \bra{\psi}\OO(t)\ket{\psi}  = \frac{1}{\mathcal{Z}} \Tr{\OO e^{-\beta H}}
\end{align}
where the inverse temperature $\beta$ is fixed by the energy density in the state $\ket{\psi}$. 
In a many-body localized phase, local observables fail to reach their thermal values and Eq.~\eqref{eq:thermaldefn} does not hold. 

In the opposite order of limits from Eq.~\eqref{Eq:LimitOrder}, the expectation values are controlled by the eigenstates $\ket{E}$.
For example,
\begin{align}
	\lim_{t\to\infty} \frac{1}{t} \int_0^t dt' \bra{\psi}\OO(t')\ket{\psi} =  \sum_{E} |\langle \psi| E\rangle |^2 \bra{E}\OO\ket{E}
\end{align}
There are two standard ans\"{a}tze for the structure of many-body eigenstates: 1) the \emph{eigenstate thermalization hypothesis} (ETH) and 2) l-bits. 
The first is generally associated with thermalization and the second with MBL at all energy densities or full MBL
\footnote{We do not consider the possibility of many-body mobility edges in this paper.}.
One of the central points of this article is that this association is incomplete.
Below we review the two ans\"{a}tze in more technical detail.

For concreteness, we consider a system of $V = L^d$ spin $1/2$ degrees of freedom $\sigma_{\mathbf{i}}$ on a $d$-dimensional lattice. 
The Hamiltonian $H$ is short-ranged with random local couplings and traceless (so that $E=0$ corresponds to infinite temperature). 

\subsection{The l-bit ansatz}
An l-bit (localized bit) is defined by a quasi-local Pauli operator $\tau_\mathbf{i}^z$ localized near site $\mathbf{i}$. 
Quasi-locality means that the expansion of $\tau$ in the physical $\sigma$ basis,
\begin{align}
	\tau_\mathbf{i}^z &= \sum_{\mathbf{j},\alpha} K^\alpha_\mathbf{ij} \sigma^\alpha_\mathbf{j} + \sum_{\mathbf{jk}, \alpha\beta} K^{\alpha\beta}_\mathbf{ijk} \sigma^\alpha_\mathbf{j}\sigma^\beta_\mathbf{k} + \cdots
\end{align}
has coefficients $K^{\alpha_1\cdots\alpha_m}_{\mathbf{i} \mathbf{j}_1\cdots \mathbf{j}_m}$ which typically decay exponentially with both the radius $d_{\textrm{max}}$ of the cluster $\mathbf{i}, \mathbf{j}_1 \cdots \mathbf{j}_m$ and the Hamming weight $m$:
\begin{equation}
K^{\alpha_1\cdots\alpha_m}_{\mathbf{i} \mathbf{j}_1\cdots \mathbf{j}_m} \sim \exp{\left(-\frac{ d_{\textrm{max}}}{\tilde{\xi}}- \frac{m}{\tilde{\theta}}\right )}.
\end{equation}
$\tilde{\xi}$ is a localization length and $\tilde{\theta}$ may be viewed as a Hamming localization length \footnote{In $d=1$ and for line-like operators in $d>1$, there is evidence that $\tilde{\theta}=\infty$ \cite{Khemani:aa, ros2015integrals}}. 

The l-bit ansatz posits that there exists a complete set of mutually commuting l-bits that are constants of motion \cite{huse2014phenomenology,serbyn2013local}.
These l-bits completely diagonalize the Hamiltonian:
\begin{align}
\label{Eq:MBLHamDiagonal}
H = \sum_{\mathbf{i}} J^{(1)}_{\mathbf{i}} \tau^z_{\mathbf{i}} + \sum_{\mathbf{i},\mathbf{j}} J^{(2)}_{\mathbf{i} \mathbf{j}} \tau^z_{\mathbf{i}} \tau^z_{\mathbf{j}} + \sum_{\mathbf{i}, \mathbf{j}, \mathbf{k}} J^{(3)}_{\mathbf{i} \mathbf{j} \mathbf{k}} \tau^z_{\mathbf{i}} \tau^z_{\mathbf{j}} \tau^z_{\mathbf{k}} + \dots
\end{align}
Here, the coefficients $J^{(m)}_{\mathbf{i}_1\cdots \mathbf{i}_m}$ decay analogously to the $K$ coefficients above. L-bits have been proven to exist at strong disorder in $d=1$ \cite{imbrie2014many}. 

As the l-bits are conserved, local memory of initial conditions persists as $t\to \infty$ even at finite size $L$.
Thus, the dynamics generated by Eq.~\eqref{Eq:MBLHamDiagonal} leads to MBL on taking the thermodynamic limit.

The properties of the eigenspectrum follow from the l-bit structure.
The eigenstates are completely specified by their $\pm 1$ eigenvalues under the $\tau^z$ operators: $|\{\tau\}\rangle$.
Consequently, eigenstates with the same energy density can have different eigenvalues under $\tau^z$, and can be distinguished by local measurements.
Adjacent states in the energy spectrum typically differ by an extensive number of l-bit flips.
Thus, they do not repel on the scale of mean level spacing, so that the level statistics of the eigenvalues is Poisson distributed.
Finally, the bipartite entanglement entropy of eigenstates obeys an area-law \cite{pal2010mb,Bauer:2013jw}.
Intuitively, this is because a partition only disrupts the $\tau^z$ eigenvalues straddling its boundary. 
These properties have been used in previous studies to diagnose MBL \cite{pal2010mb,kjall2014many,luitz2015many,Chen:2015aa,Baygan:2015aa,Devakul:2015aa,Baldwin:2016aa,Serbyn:2016aa}.

\subsection{The eigenstate thermalization hypothesis}
The eigenstate thermalization hypothesis for $H$ posits that the extreme limit of the microcanonical ensemble defined by a single eigenstate is locally indistinguishable from the appropriate canonical ensemble \cite{deutsch1991quantum,srednicki1994chaos,Tasaki:1998aa,Rigol:2008bh}.
Further, the matrix elements of any local operator $\mathcal{O}$ between eigenstates $\ket{E_\alpha}$ in a small energy window follow random matrix theory. 
Mathematically:
\begin{equation}
\label{Eq:ETHDefinition}
\langle E_\beta | \mathcal{O} | E_\alpha \rangle = \mathcal{O}_{th} \delta_{\alpha \beta} + \frac{r_{\alpha\beta}}{\sqrt{\rho(\bar{E})}} f(\bar{E}, \omega)
\end{equation}
where $\bar{E} = \frac{E_\alpha + E_\beta}{2}$ is the mean energy, $\omega = E_\beta - E_\alpha$ is the energy difference, $\rho(\bar{E})$ is the many body density of states at energy $\bar{E}$, $\mathcal{O}_{th}$ is the thermal expectation value of $\mathcal{O}$ at the same energy and $r_{\alpha \beta}$ are i.i.d Gaussian distributed with zero mean and variance one \footnote{The ETH ansatz applies to few-body observables in addition to local observables.}. 
The many-body density of states is related to the thermal entropy as $\rho(\bar{E}) \sim \exp(S(\bar{E}))$.
We have included a smooth energy dependence of the off-diagonal matrix elements on the scale of the many-body level spacing through the spectral function $f(\bar{E},\omega)$.
At fixed $\bar{E}$, $f(\bar{E},\omega)$ decays exponentially at large $\omega$. 
It is closely related to the spectral density of $\mathcal{O}$ and encodes the dynamic susceptibilities within linear response \cite{srednicki1999thermal,DAlessio:2015aa}.

Using the ETH ansatz, it is easy to show that \emph{every} initial state reaches local thermal equilibrium as $t\to\infty$, up to corrections that are exponentially small in the volume $V$. 
However, as we will see, the system need not thermalize in the thermodynamic limit if the spectral functions $f$ vanish as $L\to\infty$.

The properties of the eigenspectrum follow from Eq.~\eqref{Eq:ETHDefinition}.
Eigenstates with the same energy density are locally indistinguishable from one another and the thermal ensemble as $\langle E | \mathcal{O} |E \rangle = \mathcal{O}_{th}$ for all such states (up to exponentially small corrections in the volume $V$).
Next, as the off-diagonal matrix elements in Eq.~\eqref{Eq:ETHDefinition} are much larger than the typical level spacing, we expect that the many-body spectrum exhibits level repulsion.
Further, the entanglement entropy in eigenstates coincides with the thermal entropy, and thus obeys a volume law at all finite energy densities.
All these properties differentiate the l-bit from the ETH system.

\section{Approximately conserved l$^*$-bits}
A finite sized system that locally equilibrates and satisfies ETH can nevertheless fail to do so in the thermodynamic limit if the local dynamics become sufficiently slow.
Below, we show that a system with approximately conserved quasi-local l$^*$-bits realizes this scenario.
The information about the dynamical phase of the system is hidden in the functions $f$ which vanish as $L \to \infty$.

We propose that MBL should be described by these l$^*$-bits. 
In Sec.~\ref{Sec:PhenomenologyMBLstar}, we discuss the consequences and contrast them with the fully conserved l-bit ansatz.
We then turn to the well-studied $d=1$ case and show that the l$^*$-bit ansatz permits an eigenstate phase transition as a function of energy density from ETH violating to ETH satisfying states.
This leads to the phase diagram in Fig.~\ref{Fig:PhaseDiagrams}, in which the MBL system is described by l-bits at strong disorder and l$^*$-bits at intermediate disorder, and the dynamical and eigenstate transition do not coincide. 

\subsection{The l$^*$-bit}
\label{Sec:TheLstarbit}

\begin{figure}[tbp]
\begin{center}
\includegraphics[width=0.6\columnwidth]{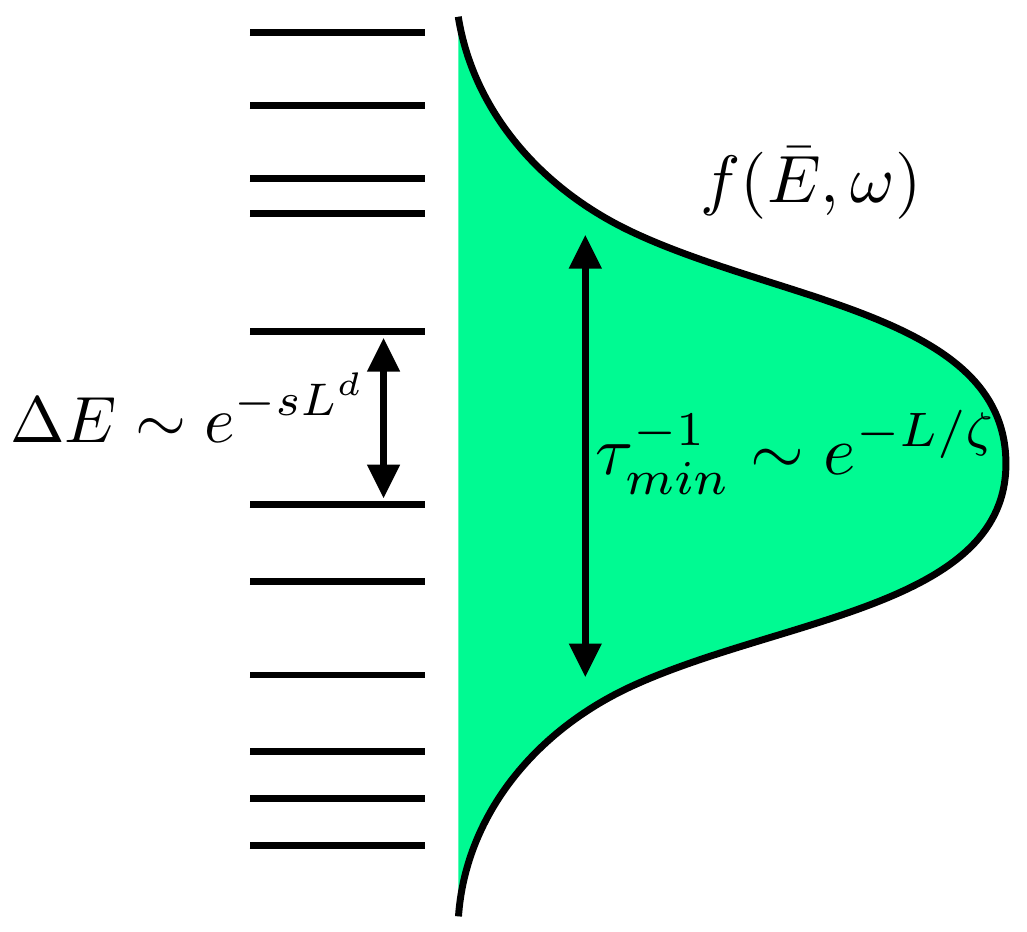}
\caption{Schematic diagram showing the $f$ function in Eq.~\eqref{Eq:ETHDefinition} for a l$^*$-bit system which satisfies ETH in $d>1$. The $f$ function is smooth on the scale of the many-body level spacing $\Delta E$, while vanishing on the much longer frequency scale $\tau_{min}^{-1}$. }
\label{Fig:ffunction}
\end{center}
\end{figure}

A l$^*$-bit is a quasi-local operator $\tau^{*z}_{\mathbf{i}}$ localized in the vicinity of site $\mathbf{i}$ that approximately commutes with the Hamiltonian. 
The norm of the commutator with the Hamiltonian is a random number that typically vanishes exponentially with $L$ 
\footnote{A random quantity $A$ which \emph{typically} vanishes exponentially has a probability distribution which satisfies $P(A> C e^{-L/\zeta})\to 0$ when $L\to\infty$ for some $C > 0$ and minimum length $\zeta$.}:
\begin{align}
\label{Eq:Lstarbit}
\|[H, \tau^{*z}_{\mathbf{i}}]\| \sim K \exp(-L/\zeta)
\end{align} 
where $\| \cdot \|$ denotes the operator norm, $\zeta$ is a localization length and $K$ has units of energy and depends subexponentially on $L$.
For simplicity, we assume $\|\tau_{\mathbf{i}}^{*z} \|=1$ below. 

Many dynamical consequences follow from the definition in Eq.~\eqref{Eq:Lstarbit}. 
Suppose the system is prepared in the state $\ket{\psi}$ at $t=0$. 
The subsequent dynamics is probed by unequal time correlators of the form:
\begin{align}
C(t) \equiv \bra{\psi} \tau^{*z}_{\mathbf{i}}(t) \mathcal{O}\ket{\psi}
\end{align}
where $\mathcal{O}$ is some local operator (for simplicity of norm 1). 
The variation of $C(t)$ is exceedingly slow. 
Taking the time derivative of $C(t)$ and using the Heisenberg equation of motion:
\begin{align}
\frac{d C(t)}{dt} = i \bra{\psi} [H,\tau^{*z}_{\mathbf{i}}(t)] \mathcal{O} \ket{\psi}
\end{align}
This derivative is upper-bounded by: 
\begin{align}
\label{Eq:UpperBound}
\left|\frac{d C(t)}{dt}\right| \lesssim K \exp(-L/\zeta) 
\end{align}
using Eq.~\eqref{Eq:Lstarbit} and the sub-multiplicative property of the norm. 
Thus, \emph{the minimum time} required for $C(t)$ to change by some fixed amount $\Delta C$ is:
\begin{align}
\label{Eq:Taumin}
\tau_{min} = \frac{\Delta C}{K} \exp(L/\zeta)
\end{align}
Note that the system need not reach a steady state (thermal or otherwise) on the time scale $\tau_{min}$; Eq.~\eqref{Eq:Taumin} is merely the minimum time needed to do so (with probability going to 1).

As $\tau_{min}$ diverges as $L \to \infty$, the l$^*$-bit becomes constant in the thermodynamic limit.
That is, $C(t) = C(0)$ for all times and 
\begin{align}
\label{Eq:Dynamicalphaseorder}
\lim_{t \to \infty}\lim_{L \to \infty}  C(t) = C(0)
\end{align}
Generic local operators in the vicinity of $\mathbf{i}$ overlap $\tau^{*z}_\mathbf{i}$ and similarly fail to reach their thermal values for any state $|\psi \rangle$.

At finite system size, however, the l$^*$-bit is not conserved and nothing prevents it from decaying to its thermal value,
\begin{align}
\label{Eq:FiniteSizeOrder}
\lim_{L \to \infty} \lim_{t \to \infty} C(t) = C_{th}
\end{align}
where $C_{th}$ is the disconnected expectation value in the appropriate Gibbs ensemble:
\begin{align}
 C_{th} =\left( \frac{1}{\mathcal{Z}} \Tr \tau^{*z}_{\mathbf{i}} e^{-\beta H} \right)\left(\frac{1}{\mathcal{Z}} \Tr \mathcal{O} e^{-\beta H} \right)
\end{align}
Indeed, ETH is perfectly consistent with the existence of l$^*$-bits for suitable functions $f$.
As a system that satisfies ETH thermalizes at finite size, this allows for Eq.~\eqref{Eq:Dynamicalphaseorder} and Eq.~\eqref{Eq:FiniteSizeOrder} to be satisfied simultaneously.

In a bit more detail, let us suppose that the system satisfies ETH at finite size $L$ but also contains an l$^*$-bit $\lstar$. 
To translate the constraint imposed by the l$^*$-bit Eq.~\eqref{Eq:Lstarbit} to functions defined by ETH, consider the correlation function:
\begin{align}
	\mathcal{I} = \bra{E}  [H,\lstar]^2 \ket{E}
\end{align}
in the eigenstate $\ket{E}$.
Using the ETH ansatz, it is straightforward to show that $\mathcal{I}$ is related to the spectral function $f$ of $\lstar$ as:
\begin{align}
	\mathcal{I} = \int_{-\infty}^\infty d\omega\, e^{-\beta \omega/2}\omega^2 |f(E,\omega)|^2
\end{align}
so long as $f$ is smooth on the scale of the many-body level spacing.
For completeness, we include the derivation in Appendix~\ref{App:ETHDerivation}.
Using Eq.~\eqref{Eq:Lstarbit}, 
\begin{align}
\int_{-\infty}^\infty d\omega\, e^{-\beta \omega/2}\omega^2 |f(E,\omega)|^2 &\lesssim K^2 \exp(-2L/\zeta) \label{Eq:Omega2SumRule}
\end{align}

As $\lstar$ is a quasi-local operator, its connected correlator is $O(1)$ in the eigenstate:
\begin{align}
\bra{E} \lstar \lstar \ket{E} - \bra{E} \lstar \ket{E}^2 \sim O(1),
\end{align}
This imposes a second sum rule on the spectral function:
\begin{align}
\label{Eq:ProbDensity1}
\int d\omega e^{-\beta \omega/2} |f(E,\omega)|^2 \sim O(1)
\end{align}
The derivation follows the same steps as Appendix~\ref{App:ETHDerivation}.

Eq.~\eqref{Eq:ProbDensity1} implies that $e^{-\beta \omega/2} |f(E,\omega)|^2$ is proportional to a probability density.
The first sum rule, Eq.~\eqref{Eq:Omega2SumRule}, then forces the weight in the distribution to concentrate around $\omega=0$ on a scale that vanishes atleast as quickly as $\tau_{min}^{-1}\sim \exp(-L/\zeta)$ (Fig.~\ref{Fig:ffunction}).
In $d>1$, $f$ can be smooth on the scale of the many body level spacing $\Delta E \sim \exp(-s L^d)$ at entropy density $s$, while showing variation on the much longer scale $\tau_{min}^{-1}$:
\begin{align}
\tau_{min}^{-1} \sim \exp(-L/\zeta), &\quad \Delta E\sim \exp(-sL^d) \\
\Rightarrow \tau_{min}^{-1} \gg \Delta E &\quad \textrm{as } L\to \infty, d>1
\end{align}
Thus, ETH and l$^*$-bits are perfectly consistent for suitable functions $f$ at all energies $E$. 

We end with a few comments. 
In the l-bit model of Ref.~\cite{huse2014phenomenology}, it is argued that the effective decay of two-l-bit interactions is controlled by an energy dependent localization length \footnote{We thank D. Huse for drawing our attention to the energy dependence of $\zeta$.}. 
This arises due to correlations in the higher-body couplings $J$ in Eq.~\eqref{Eq:MBLHamDiagonal} rather than any explicit energy dependence in the typical decay of couplings. 
Similarly, it seems likely that an effective energy dependent localization length $\zeta(e)$ can be defined, for example by projecting the commutator in Eq.~\eqref{Eq:Lstarbit} into energy windows. 
In the analysis of the boundary instability in Sec.~\ref{Sec:BoundaryInstab} for $d>1$, it is clear that $\zeta(e)$ coincides with the effective decay length defined in Ref.~\cite{huse2014phenomenology}. 
The energy dependence of $\zeta$ does not qualitatively modify the argument for the coexistence of ETH and l$^*$-bits in $d>1$ and for simplicity we will ignore it henceforth.

Next, $d=1$ is clearly special as $\tau_{min}^{-1}$ and the level spacing compete.
Thus, the above argument only applies at energy density $e=E/V$ if:
\begin{align}
\label{Eq:Conditiond1}
\zeta(e) s(e) >1 
\end{align}
where we have explicitly indicated the energy dependence.
We discuss the implications on the phase diagram in the next section. 
Next, the definition of the l$^*$-bit assumes exponential spatial localization of the operator and exponential suppression of the commutator in system size. 
It is clear that it could be generalized to other kinds of decays with $L$.
Finally, we note that a system with a single l$^*$-bit is not likely to be robust, because a generic local perturbation would mix the l$^*$-bit with other operators that are not conserved.

\subsection{Phenomenology of MBL with l$^*$-bits ($d>1$)}  
\label{Sec:PhenomenologyMBLstar}

\begin{center}
  \begin{table*}
        \caption{Properties of systems that are thermal, have l-bits or have l$^*$-bits}
        \label{Tab:PropSummary}
        \begin{tabular}{|l|l|l| l |}
        \hline
        & Thermal & l-bits  & l$^*$-bits \\
       \hline
       Eigenstate thermalization hypothesis (ETH) & Yes & No & Yes \\
       \hline
       Eigenstate entanglement entropy & Volume & Area & Volume \\
       \hline
       Level repulsion & Yes & No & Yes \\
       \hline
       Local equilibration time & Poly($L$) & Infinite & Exp($L$) \\
       \hline
       Forbidden eigenstate orders & No & Yes & No\\
       \hline
       Dynamical phase at all energy densities & Thermal & MBL & MBL \\
       \hline   
        \end{tabular}
    \end{table*}
\end{center}

The discussion above suggests that a more general phenomenology of full MBL is provided by a collection of $N = L^d - O(L^{d-1})$ algebraically independent l$^*$-bits. 
This relaxes the l-bit construction in two ways. 
First, the localized bits are only approximately conserved at finite size. Second, as we will see in Sec.~\ref{Sec:BoundaryInstab}, this accommodates the inclusion of thermal layers. 
Further, we assume that the eigenstates are as random as possible subject to the constraint in Eq.~\eqref{Eq:UpperBound} -- i.e. they satisfy the ETH ansatz with $f$ appropriately vanishing in the thermodynamic limit.
Note that l-bits and l$^*$-bits lead to MBL at all energy densities due to their definition in terms of operator norms.

We summarize the properties of the MBL system with l$^*$-bits in Table~\ref{Tab:PropSummary}  for $d>1$, and contrast it with a MBL system with conserved l-bits and a thermalizing ETH system. 
Many of the properties follow immediately from the previous discussion. 
A few warrant further consideration.

The scaling of the eigenstate entanglement entropy $S_E$ with subregion size is a commonly used measure of MBL \cite{kjall2014many,luitz2015many,Baygan:2015aa,Devakul:2015aa,Grover:2014aa}.
With l-bits, $S_E$ scales with an area law, while in ETH systems, $S_E$ coincides with the thermal entropy and accordingly exhibits volume law scaling.
As the eigenstates in the l$^*$-bit system satisfy ETH, $S_E$ shows a volume law despite being MBL in the thermodynamic limit.
The special structure of $f$ in the l$^*$-bit system could be captured by a sub-leading term in $S_E$ at $O(L)$; we leave this question for future study. 

Level repulsion arises when the off-diagonal matrix elements of local operators between adjacent eigenstates are much larger than the level spacing. 
In thermal and l$^*$-bit systems, by the ETH ansatz, the former $\sim e^{-s L^d / 2}$ are much larger than the latter $\sim e^{-s L^d}$. 
The matrix elements are only enhanced by the exponentially in $L$ divergent $f(\bar{E},0)$ in the l$^*$-bit system.
Thus, the levels repel.
In the l-bit system, on the other hand, typical matrix elements $\sim e^{-L^{d}/2\theta}$ have to be much less than the level spacing $\sim e^{-s L^d}$ in order for the l-bits to be stable. 
See Ref.~\cite{Serbyn:2015aa} for more details.

By assumption, l$^*$-bits change on a time scale longer than $\tau_{min}\sim e^{O(L)}$.
In thermalizing systems, on the other hand, the slowest modes are typically diffusive and the equilibration time for local operators is at most polynomially large in $L$.
Moreover, the short time dynamics of any local operator would still take place on $O(1)$ timescale. 
This differentiates thermal systems from l$^*$-bit systems at finite size. 

Finally, the seminal work of Huse et al \cite{Huse:2013aa} proposed that MBL could protect long-range order at finite energy densities even when forbidden by equilibrium statistical mechanics. 
They argued that the order manifests in individual eigenstates and sub-classified MBL phases according to their eigenstate orders.
Several works have since extended these classifications \cite{Bauer:2013jw,Chandran:2014aa,Bahri:2015aa,Potter:2015aa,von-Keyserlingk:2016aa,Potter:2016aa,Else:2016aa}. 
All these classifications rely on the identification of l-bits. 
In the presence of l$^*$-bits, it is still possible to have dynamically frozen order in the thermodynamic limit. 
This order would however not show up as an eigenstate order.

\subsection{Phenomenology of MBL with l$^*$-bits ($d=1$)}
As mentioned at the end of Sec.~\ref{Sec:TheLstarbit}, $d=1$ is special because $\tau_{min}^{-1}$ and the many-body level spacing compete (see Eq.~\eqref{Eq:Conditiond1}).
There are also rigorous results at strong disorder about the existence of a complete set of l-bits at finite size \cite{imbrie2014many}.
The l-bits impose $\zeta=0$ and Eq.~\eqref{Eq:Omega2SumRule} cannot be satisfied by a smooth $f$ function.

As the strength of disorder is reduced, $\zeta$ could become non-zero producing an intermediate l$^*$-bit regime. 
In this section, we will explore the consequences and the existing evidence for this scenario.
The general mechanism for producing l$^*$-bits in $d=1$ is an interesting open question; the instability identified in  Sec.~\ref{Sec:BoundaryInstab} only applies in $d>1$. 

\subsubsection{ Apparent many-body mobility edges}
An outstanding theoretical problem regards the existence and description of many-body mobility edges \cite{De-Roeck:2016aa, laumann2014many, Baldwin:2016aa, kjall2014many, Mondragon-Shem:2015aa, luitz2015many} --- that is, delocalization transitions as a function of energy density within a single sample. 
Neither the l-bit nor the l$^*$-bit ansatz permit mobility edges as they lead to MBL at all energy densities in the thermodynamic limit. 
However, in $d=1$, there can be an eigenstate phase transition within the l$^*$-bit phase between ETH-satisfying and ETH-violating states when $\zeta s = 1$. 
This is shown as the red line in Fig.~\ref{Fig:PhaseDiagrams} which lies entirely within the MBL phase.
This transition need not coincide with the true dynamical mobility edge in the thermodynamic limit, shown as a black line in Fig.~\ref{Fig:PhaseDiagrams}.

Eigenstate based studies could incorrectly identify the red line as the mobility edge. 
A physical diagnostic of the true MBL transition is provided by the relaxation time of local observables after global quenches. 
The l$^*$-bit phase would exhibit an exponentially growing relaxation time with system size even when the eigenstates satisfy ETH. 
In contrast, in the thermal phase, relaxation is limited by (sub-)diffusion which produces power laws in system size. 
In the well-studied random field Heisenberg chain, there is some evidence that the transition identified from the variance of late time observables after global quenches \cite{Singh:2016aa} lies at a lower disorder value than the eigenstate phase transition \cite{luitz2015many}, consistent with the phase diagram of Fig.~\ref{Fig:PhaseDiagrams}.
This is worthy of further investigation.

\subsubsection{Inverse participation ratios}
The l$^*$-bit ansatz suggests an numerically accessible measure to detect the delocalized eigenstates in the pink region in Fig.~\ref{Fig:PhaseDiagrams}.
Consider $e=0$ for simplicity. 
The inverse participation ratio (IPR) in the energy basis of an eigenstate acted on by a local operator $\mathcal{O}\ket{E_\alpha}$ typically decays as:
\begin{align}
I = \sum_{\beta} |\langle E_{\beta} | \mathcal{O} | E_\alpha| \rangle|^4 \sim 2^{-L\eta}.
 \end{align}
For an ETH system without l$^*$-bits, $I \sim 2^{-L}$ from Eq.~\eqref{Eq:ETHDefinition}.
For the l$^*$-bit system on the other hand,  $I  \sim 2^{-L} \tau_{min} \sim 2^{-L} e^{+L/\zeta}$, so that $\eta<1$.
Previous numerical studies \cite{de2013ergodicity,luitz2015many,Baldwin:2016aa} have looked at other (Hilbert space) IPRs but we expect the IPR defined above to be more sensitive to l$^*$-bit structure of the eigenstates.

\section{Boundary instability of l-bit systems} 
\label{Sec:BoundaryInstab}

\begin{figure}[tbp]
\begin{center}
\includegraphics[width=\columnwidth]{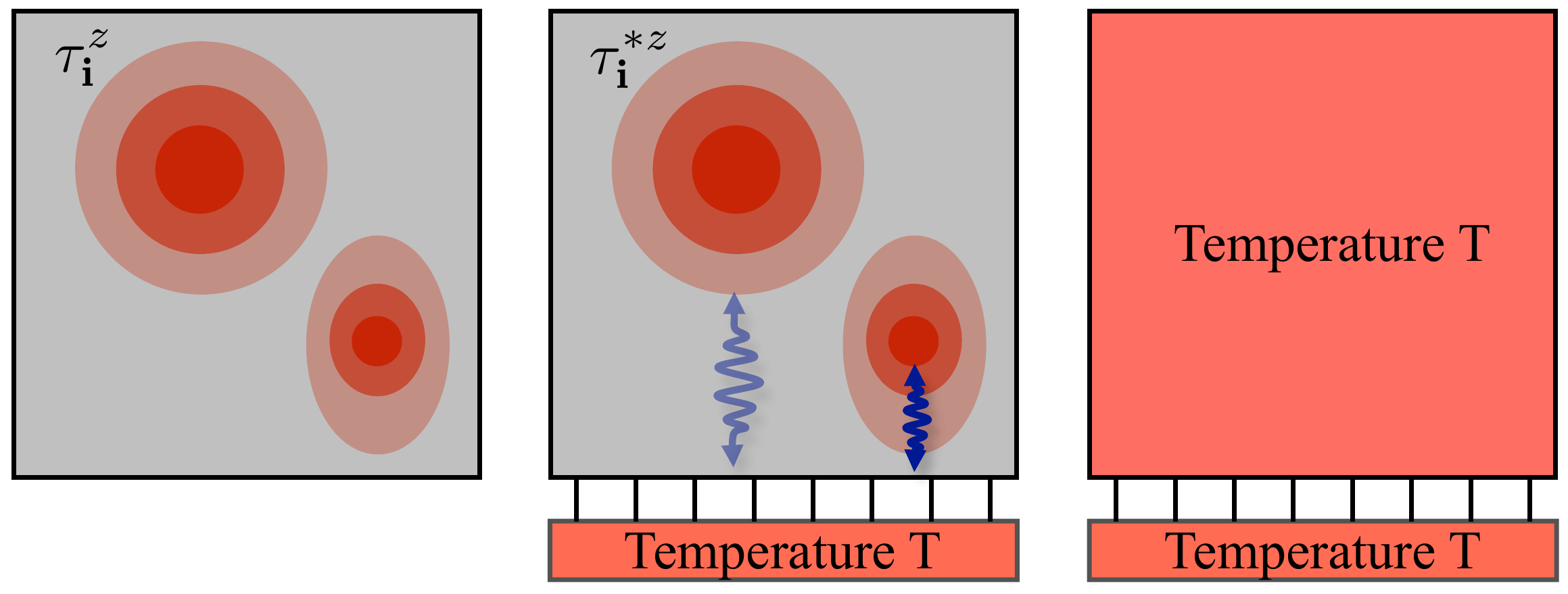}
\caption{Left: A MBL system with l-bits $\tau_{\mathbf{i}}^z$. The shading indicates that the operators are quasi-local with exponentially decaying weights away from the localization center. Middle: The l-bit system coupled to a thermal boundary layer at temperature $T$. The boundary induces incoherent l-bit flips in the bulk with a rate that decreases exponentially with the distance from the boundary. Right: As $t\to \infty$ at finite size, the bulk and boundary are locally thermal at temperature $T$. }
\label{Fig:Lstarbits}
\end{center}
\end{figure}

In this section, we show that l-bits are unstable due to their extreme sensitivity to boundary effects.
Specifically, we argue that a thin thermal boundary layer at the surface is sufficient to cause all local operators to decay on an exponentially divergent time scale.
The original l-bits remain long-lived only in the thermodynamic limit, becoming the l$^*$-bits discussed in the previous section. 
Our arguments do not apply in $d=1$ where boundary regions are finite.
It turns out that the $d=2$ case is marginal as we describe at the end of the section.

\subsection{Long-time Thermalization}

Consider a finite, $d$-dimensional l-bit system of linear dimension $L$ with Hamiltonian $\Hbulk$. 
This system is coupled to a $(d-1)$-dimensional boundary layer with spin 1/2 degrees of freedom $\gamma_{\mathbf{i}}$ and Hamiltonian $\Hbdy$ that satisfies the ETH ansatz (see Fig.~\ref{Fig:Lstarbits}).
We assume that the coupling is local and only connects the physical spins $\sigma$ at the edge of the system to the boundary. For example,
\begin{equation}
\label{Eq:Hint}
\Hint = \lambda \sum_{\mathbf{i} \in \text{edge}} \sigma^{x}_{L,\mathbf{i}} \gamma^{x}_{\mathbf{i}} 
\end{equation}
$\lambda$ characterizes the strength of the interaction between the system and the boundary.
In terms of the $\tau$ operators, $H_{int}$ contains $\tau^{x,y,z}$ operators with weights decaying exponentially into the bulk:
\begin{equation}
\label{Eq:HintQLocal}
\Hint = \lambda \sum_{\mathbf{i} \in \text{edge}} \left(\sum_{\mathbf{j}} C_{\mathbf{ij}} \tau^{x}_{\mathbf{j}} + \ldots \right) \gamma^{x}_{\mathbf{i}} 
\end{equation}
where $C_{\mathbf{ij}} \sim \exp(-R_{\mathbf{ij}}/\xi)$, $R_{\mathbf{ij}}$ is the distance between the site $(L,\mathbf{i})$ and $\mathbf{j}$ and we have not written out the other terms in the quasi-local expansion.
For convenience, we assume $\Hbulk, \Hbdy, \Hint$ are traceless.

In the absence of the coupling ($\lambda=0$), the eigenstates of the system are given by $\ket{E, \{\tau\}}$ where $E$ specifies an eigenenergy of the boundary and $\{\tau\}$ is an l-bit configuration in the bulk.
The $\tau^x \gamma^x$ terms in $H_{int}$ induce single l-bit flips in the bulk and transitions between eigenstates in the boundary. 
To leading order in $\lambda$, the total transition rate from $|E,\{\tau\}\rangle $ to all states $|E',\{\tau'\}\rangle $ in which only the $m$th bit is flipped ($\tau'_{\textbf{m}} = -\tau_{\textbf{m}}$) is given by Fermi's Golden Rule:
\begin{align}
\Gamma^{\textbf{m}}_{+\rightarrow-} = \pi \lambda^2 \sum_{\mathbf{i} \in \text{edge}}  C^{2}_{\textbf{im}} e^{-\beta \Delta_{\textbf{m}}/2} |f_{\mathbf{i}}(E, -\Delta_{\textbf{m}})|^2
\end{align}
Above, $\beta$ is the inverse temperature of the boundary layer, $\Delta_{\textbf{m}}$ is the energy difference between the l-bit state with $+\tau_{\textbf{m}}$ and $-\tau_{\textbf{m}}$ and $f_{\mathbf{i}}$ is the spectral function associated with $\gamma^{x}_{\mathbf{i}}$ in the ETH ansatz. 
Note that the expression neglects the shift in the energy of the bath $E$ as $\Delta_{\textbf{m}} \sim O(1) \ll E \sim O(L^{d-1})$. 
As $C_{\mathbf{im}}$ decays exponentially with $R_{\mathbf{im}}$, the transition rate is dominated by the term with the smallest $R_{\mathbf{im}}$.
Denoting the distance to the boundary by $R$,
\begin{align}
\label{Eq:FGR}
\Gamma^{\textbf{m}}_{+\rightarrow-} \sim  \pi \lambda^2   e^{-2R/\xi} e^{-\beta \Delta_{\textbf{m}}/2} |f(E, -\Delta_{\textbf{m}})|^2
\end{align}

The Fermi's Golden Rule calculation assumes that the boundary layer is `large', so that its many-body level spacing is not visible to the l-bits. 
More quantitatively, it is only valid when the rates $\Gamma^{\textbf{m}}_{+\rightarrow-}$ far exceed the many-body level spacing on the boundary $\Delta E$.
The smallest transition rates are associated with the l-bits deep in the bulk at a distance $R \sim \alpha  L $ away from the boundary:
\begin{align}
\label{Eq:GammaMin}
\Gamma_{min} \sim e^{-2\alpha L/\xi}
\end{align}
In $d>2$, these exceed the many-body level spacing on the boundary:
\begin{align}
\Gamma_{min} \gg \Delta E \sim e^{-sL^{d-1}} \quad d>2
\end{align}
with the inequality getting better at larger $L$. 
Above $s$ is the boundary entropy density at energy $E$.
Thus, at weak coupling $\lambda$, the perturbative estimate of the decay rate in Eq.~\eqref{Eq:FGR} is trustworthy for all bulks l-bits in $d>2$.

In $d>2$, the boundary layer induces single l-bit flip transitions everywhere in the bulk.
Further, the transition rates satisfy detailed balance:
\begin{align}
\frac{\Gamma_{+\rightarrow-}}{\Gamma_{-\rightarrow+}} = e^{-\beta \Delta}
\end{align}
where we have suppressed the index $\mathbf{m}$ for clarity.
This follows from the symmetric property of the $f$ function: $|f(E, -\Delta_{\textbf{m}})|^2 = |f(E, \Delta_{\textbf{m}})|^2$.
Any $f$ function associated with a local operator $\mathcal{O}$ must be symmetric as $|\bra{E_\beta} \mathcal{O} \ket{E_\alpha}|^2 = |\bra{E_\alpha} \mathcal{O} \ket{E_\beta}|^2$.

If we assume that the induced dynamics is Markovian, then detailed balance guarantees the equilibrium distribution of the bulk is the Boltzmann-Gibbs distribution at the same temperature as the boundary.
Thus, a thermalizing ETH boundary layer destabilizes a bulk l-bit system in $d>2$ and leads to local thermal expectation values at infinite time.
This does not mean that the combined system thermalizes in the thermodynamic limit as the time-scale for single l-bit flips diverges exponentially with $L$.
The system is therefore MBL and is rightly described by l$^*$-bits, as we discuss next.

We end this subsection with several comments. 
First, higher order terms in the quasi-local expansion of $H_{int}$ (Eq.~\eqref{Eq:HintQLocal}) lead to multi l-bit flip processes at leading order in $\lambda$.
As the transitions are incoherent, we expect these terms to enhance local thermalization but continue to satisfy detailed balance.
Second, the single spin flip decay rate in Eq.~\eqref{Eq:FGR} neglects the shift in the boundary energy. 
This is clearly a good approximation for the initial decays. 
As the bulk system explores its phase space, it typically needs to absorb $O(L^{d/2})$ energy from the bath. 
For $d>2$, this energy is boundary subextensive and it is consistent to neglect it. 
Finally, as the stationary states are thermal, we expect the eigenstates themselves to be thermal.
Within the Markov approximation, the joint eigenstates can be represented with Lorentzian weights in $\ket{E, \{\tau\}}$.

The marginal case for the above argument is $d=2$. 
First, the rate in Eq.~\eqref{Eq:FGR} and the level spacing on the boundary $e^{-s L}$ can compete. 
For strong enough disorder in the bulk (small enough localization length $\xi$), the individual l-bits further than $R = \xi s L/2$ from the boundary fail to decay to leading order in $\lambda$. 
Moreover, even in the case where $R > L$ and all l-bits can decay, the typical energy fluctuations $O(L^{d/2}) = O(L)$ in the bulk in the presumed Markov equilibrium are of order the energy in the boundary $O(L)$. 
This implies that the Markov assumption cannot hold.

\subsection{Emergence of l$^*$-bits}

At finite $\lambda$, the original l-bits deep in the bulk become l$^*$-bits.
Consider,
\begin{equation}
[H,\tau^z_{\mathbf{i}}]=[H_b+H_\partial+H_{int},\tau^z_{\mathbf{i}}]=[H_{int},\tau^z_{\mathbf{i}}].
\end{equation}
$H_{int}$ only involves the physical bits on the boundary. 
As the expansion of each physical bit in terms of the l-bits is quasi-local, 
\begin{equation}
\|[H_{int},\tau^z_{\mathbf{i}}]\| \sim \lambda e^{-R/\zeta'},
\end{equation}
where $\zeta'$ is a localization length and we have neglected a polynomial prefactor in $L$. Thus, $\tau^z_{\mathbf{i}}$ is an l$^*$-bit if $R \sim O(L)$. Further, the number of independent l$^*$-bits $N$ scales as $L^d$ as $L\to \infty$, as required in Sec.~\ref{Sec:PhenomenologyMBLstar}. 
Note that the scaling is only asymptotic as $N < L^d$ at any finite size $L$. 
This is because $\tau$ bits at a finite distance away from the boundary always have a finite lifetime and are not conserved as $L\to\infty$. 

Finally, we comment on the connection with the perturbative calculation of l-bits presented in Ref.~\cite{ros2015integrals}, which is valid for sufficiently strong  statistically homogenous disorder in the thermodynamic limit.
At finite size, the same method produces l$^*$-bits at distance $r \sim L$ if one stops the real-space perturbation theory when the support of the operator reaches the boundary.  
The commutator with $H$ would then be $g^{O(L)}$ where $g$ is the bulk coupling. 
Continuing with the perturbation theory including operators on a thermal boundary layer would cause the series to diverge. 
This suggests that l-bits do not exist in this system while the l$^*$-bits do.

\section{Experiments}
\label{Sec:Experiments}

\begin{figure*}[tbp]
\begin{center}
\includegraphics[width=2\columnwidth]{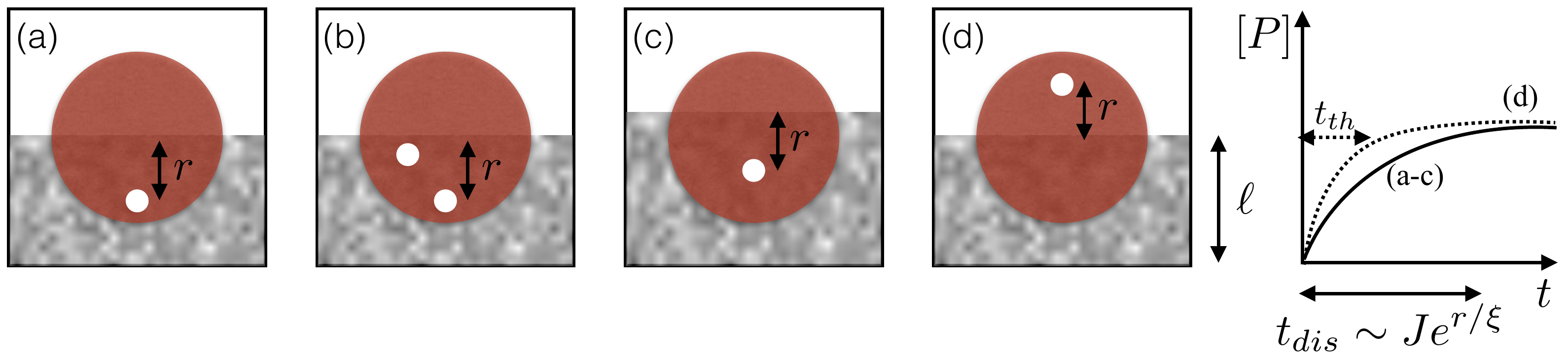}
\caption{ (a-d) Real space pictures of potential ultracold atomic experiments to probe the effect of thermal layers on localized bulk. Red indicates the cloud of atoms and grey the speckle potential applied to only part of the cloud. White spots indicate holes in the initial density.
(e) The mean parity of density in the hole as a function of time in the experiments indicated in (a-d).
 }
\label{Fig:ExperimentalPlot}
\end{center}
\end{figure*}

In this section, we describe experiments in current cold atomic setups that directly test the l$^*$-bit phenomenology of MBL.
With the observation of MBL in two dimensions using a quantum-gas microscope \cite{Choi:2016aa}, such experiments are well within the scope of current technology. 
The single-site resolution for imaging and manipulating hyperfine states of atoms in an optical lattice \cite{Bakr:2009aa,Bakr547,Sherson:2010aa} allows to study the local deviations from thermal equilibrium. 
This has been exploited to probe the diverging length scale at the dynamical transition between MBL and thermal phases in two dimensions \cite{Choi:2016aa}. 

Briefly, the current experiments prepare an initial Mott insulating state of bosonic Rubidium-87 atoms with a spatially varying density profile in the $x-y$ plane. 
Initially, all the atoms occupy the left half of the harmonic trap at an average filling of one per site, resulting in a sharp domain wall at $x=0$. 
On release, the atoms expand into the right half of the trap in the background of on-site disorder that is varied from run to run.
In the thermal phase, the cloud reaches a steady state with uniform density (up to trap effects), while in the localized phase, the memory of the inhomogeneity in the initial density profile persists at late times. 
The authors measure the parity of the atoms on each site at different times and characterize the two phases according the long time imbalance in the number of atoms between the right and left halves.

The geometry of a localized bulk connected to a thermal boundary is easily accessible in the current experimental setup as the on-site disorder of the atoms can be spatially varied at will and can be experimentally characterized.
The simplest possibility is to create a domain wall in the on-site disorder in each run, as shown in Fig.~\ref{Fig:ExperimentalPlot} (a-d) in gray scale.
In these plots, the on-site disorder is drawn from a uniform uncorrelated distribution in the bottom $\ell$ rows and is absent in the top rows.
The disorder-free region thus functions as the thermal region, while the strength of the disorder is chosen to be sufficiently large so that the bottom $\ell$ rows are localized.

In order to characterize the dynamics of the combined system, we propose using the single-site control in the experiments to create density holes in the trapped cloud at $t=0$ and track their parity in time. In Fig.~\ref{Fig:ExperimentalPlot} (a-d), the cloud of atoms in the trap is shown in solid red, while the density hole is shown in white. In Fig.~\ref{Fig:ExperimentalPlot}(a) for example, the hole is created at a distance $r$ from the domain wall in the disorder distribution in the localized region. The setup is flexible and offers many experimental knobs to test different aspect of the dynamics. The experimental knobs include the distance $r$, the strength of the disorder, the number of density holes (b), the size of the localized region $\ell$ (c) and the environment of the density hole (d). 

The discussion in Sec.~\ref{Sec:BoundaryInstab} suggests how the density hole relaxes  in time. Consider first a hole created in the disordered region. If the localized region is described only by l$*$-bits, then the relaxation time $t_{dis}$ should be exponentially sensitive to $r/\xi$, where $\xi$ is a localization length. We expect that this time is insensitive to the number of holes and to the size of the disordered region $\ell$, as long as the disorder-free region acts as a thermal bath. In contrast, in the disorder-free region that is expected to be thermal, the relaxation time $t_{th}$ should be independent of $r$ and other details of the disordered region.

The proposed experiments are interesting for many reasons. First, the discussion in Sec.~\ref{Sec:BoundaryInstab} is marginal in $d=2$. Specifically, at strong enough disorder, we expect l-bits far away from the boundary to be stable despite the coupling to the thermalizing region. However, the l-bits close to the boundary will decay; mapping out the dependence of the relaxation time with $r$, the strength of the disorder and the size of the thermalizing region would provide invaluable insight into the dynamics of the combined system. It would allow us to test the general validity of arguments like those in Sec.~\ref{Sec:BoundaryInstab} that rely on the competition between the typical matrix elements and typical level spacing alone. Next, the relaxation time $t_{dis}$ provides access to a different localization length $\xi$ than the imbalance studied in current experiments. Testing the relation between the two would shed light on the issue of multiple localization lengths in the localized phase. Finally, the set-up offers a controlled way to de-stabilize the localized phase and study its response to a small thermal reservoir, an important piece in the puzzle of the dynamical transition.

\section{Conclusions}

The properties of finite size samples have played a pivotal role in our understanding of Anderson localization beginning with the work of Thouless and the Gang of Four in the late 70s \cite{Edwards:1972aa,Thouless:1974aa,Abrahams:1979ud}. 
For example, Thouless observed that finite size conductance can be measured by the sensitivity of non-interacting eigenstates to boundary conditions and many numerical studies have since relied on the associated spectral statistics as a diagnostic tool \cite{Abrahams2010}. 
The success of this approach relies implicitly on the stability of the statistical properties of the eigenstates to the varying boundary conditions encountered as the thermodynamic limit is taken. 
Indeed, for non-interacting electrons, the analysis of Sec.~\ref{Sec:BoundaryInstab} does not lead to an eigenstate instability as the Fermi Golden Rule rate is much less than the polynomially small level spacing.

In this article, we have argued that this stability simply does not hold for interacting systems in dimensions greater than one.
Rather as the thermodynamic and late time limit fail to commute, the eigenstate properties of a many-body localized system may satisfy finite size ETH. 
We have offered a refined phenomenology of fully MBL systems in terms of approximately conserved l$^*$-bits which we believe is more robust than the current l-bit scenario.
The most striking consequence for existing studies is that the eigenstate transitions need not coincide with the localization transition.
We have also proposed experiments which can probe the boundary instability and test the l$^*$-bit scenario.

An analogy can be drawn between the theory of l$^*$-bits and that of the weakly interacting Fermi liquid.
Both have effective Hamiltonian descriptions which are diagonal in a basis of dressed operators. 
These Hamiltonians, however, neglect terms which cause the decay of the quasiparticles or l$^*$-bits. 
In the Fermi liquid case, the quasiparticle lifetime is controlled by the deviation of the single-particle energy from the Fermi surface while in the l$^*$-bit case it is controlled by finite size.
For a conceptually different analogy between MBL and Fermi liquids, see Ref.~\cite{Bera:2015aa}.

It is widely believed that Griffiths-type effects play an important role in the theory of the MBL transition, especially in $d=1$ \cite{Bar-Lev:2015aa,Agarwal:2015aa,Vosk:2015aa,Potter:2015ab,Gopalakrishnan:2016aa,Kerala-Varma:2015aa}. 
We withhold judgement regarding their role in $d>1$. 
It is worth pointing out that the boundary instability in Sec.~\ref{Sec:BoundaryInstab} relies on a thermalizing region whose size diverges. 
In statistically homogeneous systems at strong disorder, we therefore do not expect to find such large thermal subregions except perhaps at the physical boundaries.

A number of questions are raised by the possibility of l$^*$-bits and ETH MBL. 
The arguments regarding the absence of many-body mobility edges of Ref.~\cite{De-Roeck:2016aa} hinge on the association of finite-size ETH with thermal phases and are thus inconclusive in light of the thesis of this article.
Many of the classifications of `allowed' MBL quantum orders in both Floquet and Hamiltonian systems rely heavily on the l-bit model \cite{Huse:2013aa,Pekker:2014aa,Chandran:2014aa,Bahri:2015aa,Potter:2015aa,von-Keyserlingk:2016aa,Else:2016aa}. 
It would be very interesting to determine what of these orders and classifications survives on relaxing to the l$^*$-bit scenario.
The phase diagram of Fig.~\ref{Fig:PhaseDiagrams} suggests that the transitions identified in previous $d=1$ numerics can lie within the MBL phase. 
This raises questions about the identification of sub-diffusive phases and more generally how to identify the true dynamical transition.

\begin{acknowledgments}
We are grateful to D. Huse for stimulating discussions and for a careful reading of the manuscript, and to J.Y. Choi, C. Gross, R. Islam and P. Bordia for detailed discussions on the MBL experiments. 
We would also like to thank S.L. Sondhi, M. M\"uller, F. Huveneers, W. de Roeck, V. Khemani, F. Pollmann and S. Simon for fruitful discussions. 
We thank the Kavli Institute for Theoretical Physics (KITP) in Santa Barbara for their hospitality during the early stages of this work and the National Science Foundation (NSF) under Grant No. NSF PHY11-25915 for supporting KITP.
CRL acknowledges further support from the Sloan Foundation through a Sloan Research Fellowship and the NSF through Grant No. PHY-1520535. 
Research at Perimeter Institute is supported by the Government of Canada through Industry Canada and by the Province of Ontario through the Ministry of Research and Innovation.

\end{acknowledgments}

\bibliography{biblioMBL}

\begin{thebibliography}{94}
\expandafter\ifx\csname natexlab\endcsname\relax\def\natexlab#1{#1}\fi
\expandafter\ifx\csname bibnamefont\endcsname\relax
  \def\bibnamefont#1{#1}\fi
\expandafter\ifx\csname bibfnamefont\endcsname\relax
  \def\bibfnamefont#1{#1}\fi
\expandafter\ifx\csname citenamefont\endcsname\relax
  \def\citenamefont#1{#1}\fi
\expandafter\ifx\csname url\endcsname\relax
  \def\url#1{\texttt{#1}}\fi
\expandafter\ifx\csname urlprefix\endcsname\relax\def\urlprefix{URL }\fi
\providecommand{\bibinfo}[2]{#2}
\providecommand{\eprint}[2][]{\url{#2}}

\bibitem[{\citenamefont{Anderson}(1958)}]{anderson1958absence}
\bibinfo{author}{\bibfnamefont{P.}~\bibnamefont{Anderson}},
  \bibinfo{journal}{Phys. Rev.} \textbf{\bibinfo{volume}{109}},
  \bibinfo{pages}{1492} (\bibinfo{year}{1958}).

\bibitem[{\citenamefont{Basko et~al.}(2006)\citenamefont{Basko, Aleiner, and
  Altshuler}}]{basko2006metal}
\bibinfo{author}{\bibfnamefont{D.}~\bibnamefont{Basko}},
  \bibinfo{author}{\bibfnamefont{I.}~\bibnamefont{Aleiner}}, \bibnamefont{and}
  \bibinfo{author}{\bibfnamefont{B.}~\bibnamefont{Altshuler}},
  \bibinfo{journal}{Annals of physics} \textbf{\bibinfo{volume}{321}},
  \bibinfo{pages}{1126} (\bibinfo{year}{2006}).

\bibitem[{\citenamefont{Gornyi et~al.}(2005)\citenamefont{Gornyi, Mirlin, and
  Polyakov}}]{gornyi2005interacting}
\bibinfo{author}{\bibfnamefont{I.}~\bibnamefont{Gornyi}},
  \bibinfo{author}{\bibfnamefont{A.}~\bibnamefont{Mirlin}}, \bibnamefont{and}
  \bibinfo{author}{\bibfnamefont{D.}~\bibnamefont{Polyakov}},
  \bibinfo{journal}{Physical review letters} \textbf{\bibinfo{volume}{95}},
  \bibinfo{pages}{206603} (\bibinfo{year}{2005}).

\bibitem[{\citenamefont{Oganesyan and Huse}(2007)}]{oganesyan2007localization}
\bibinfo{author}{\bibfnamefont{V.}~\bibnamefont{Oganesyan}} \bibnamefont{and}
  \bibinfo{author}{\bibfnamefont{D.}~\bibnamefont{Huse}},
  \bibinfo{journal}{Physical Review B} \textbf{\bibinfo{volume}{75}},
  \bibinfo{pages}{155111} (\bibinfo{year}{2007}).

\bibitem[{\citenamefont{Nandkishore and Huse}(2015)}]{Nandkishore:2015aa}
\bibinfo{author}{\bibfnamefont{R.}~\bibnamefont{Nandkishore}} \bibnamefont{and}
  \bibinfo{author}{\bibfnamefont{D.~A.} \bibnamefont{Huse}},
  \bibinfo{journal}{Annual Review of Condensed Matter Physics}
  \textbf{\bibinfo{volume}{6}}, \bibinfo{pages}{15} (\bibinfo{year}{2015}).

\bibitem[{\citenamefont{Altman and Vosk}(2015)}]{Altman:2015aa}
\bibinfo{author}{\bibfnamefont{E.}~\bibnamefont{Altman}} \bibnamefont{and}
  \bibinfo{author}{\bibfnamefont{R.}~\bibnamefont{Vosk}},
  \bibinfo{journal}{Annual Review of Condensed Matter Physics}
  \textbf{\bibinfo{volume}{6}}, \bibinfo{pages}{383} (\bibinfo{year}{2015}).

\bibitem[{\citenamefont{Eisert et~al.}(2015)\citenamefont{Eisert, Friesdorf,
  and Gogolin}}]{Eisert:2015aa}
\bibinfo{author}{\bibfnamefont{J.}~\bibnamefont{Eisert}},
  \bibinfo{author}{\bibfnamefont{M.}~\bibnamefont{Friesdorf}},
  \bibnamefont{and} \bibinfo{author}{\bibfnamefont{C.}~\bibnamefont{Gogolin}},
  \bibinfo{journal}{Nat Phys} \textbf{\bibinfo{volume}{11}},
  \bibinfo{pages}{124} (\bibinfo{year}{2015}).

\bibitem[{\citenamefont{Monthus and Garel}(2010)}]{Monthus:2010vn}
\bibinfo{author}{\bibfnamefont{C.}~\bibnamefont{Monthus}} \bibnamefont{and}
  \bibinfo{author}{\bibfnamefont{T.}~\bibnamefont{Garel}},
  \bibinfo{journal}{Phys. Rev. B} \textbf{\bibinfo{volume}{81}},
  \bibinfo{pages}{134202} (\bibinfo{year}{2010}).

\bibitem[{\citenamefont{Vosk and Altman}(2013)}]{vosk2013dynamical}
\bibinfo{author}{\bibfnamefont{R.}~\bibnamefont{Vosk}} \bibnamefont{and}
  \bibinfo{author}{\bibfnamefont{E.}~\bibnamefont{Altman}},
  \bibinfo{journal}{Phys. Rev. Lett.} \textbf{\bibinfo{volume}{110}},
  \bibinfo{pages}{067204} (\bibinfo{year}{2013}).

\bibitem[{\citenamefont{Pekker et~al.}(2014)\citenamefont{Pekker, Refael,
  Altman, Demler, and Oganesyan}}]{Pekker:2014aa}
\bibinfo{author}{\bibfnamefont{D.}~\bibnamefont{Pekker}},
  \bibinfo{author}{\bibfnamefont{G.}~\bibnamefont{Refael}},
  \bibinfo{author}{\bibfnamefont{E.}~\bibnamefont{Altman}},
  \bibinfo{author}{\bibfnamefont{E.}~\bibnamefont{Demler}}, \bibnamefont{and}
  \bibinfo{author}{\bibfnamefont{V.}~\bibnamefont{Oganesyan}},
  \bibinfo{journal}{Phys. Rev. X} \textbf{\bibinfo{volume}{4}},
  \bibinfo{pages}{011052} (\bibinfo{year}{2014}).

\bibitem[{\citenamefont{\v{Z}nidari\v{c}
  et~al.}(2008)\citenamefont{\v{Z}nidari\v{c}, Prosen, and
  Prelov\v{s}ek}}]{znidaric2008many}
\bibinfo{author}{\bibfnamefont{M.}~\bibnamefont{\v{Z}nidari\v{c}}},
  \bibinfo{author}{\bibfnamefont{T.}~\bibnamefont{Prosen}}, \bibnamefont{and}
  \bibinfo{author}{\bibfnamefont{P.}~\bibnamefont{Prelov\v{s}ek}},
  \bibinfo{journal}{Physical Review B} \textbf{\bibinfo{volume}{77}},
  \bibinfo{pages}{64426} (\bibinfo{year}{2008}).

\bibitem[{\citenamefont{Pal and Huse}(2010)}]{pal2010mb}
\bibinfo{author}{\bibfnamefont{A.}~\bibnamefont{Pal}} \bibnamefont{and}
  \bibinfo{author}{\bibfnamefont{D.~A.} \bibnamefont{Huse}},
  \bibinfo{journal}{Phys. Rev. B} \textbf{\bibinfo{volume}{82}},
  \bibinfo{pages}{174411} (\bibinfo{year}{2010}).

\bibitem[{\citenamefont{Bardarson et~al.}(2012)\citenamefont{Bardarson,
  Pollmann, and Moore}}]{Bardason2012}
\bibinfo{author}{\bibfnamefont{J.~H.} \bibnamefont{Bardarson}},
  \bibinfo{author}{\bibfnamefont{F.}~\bibnamefont{Pollmann}}, \bibnamefont{and}
  \bibinfo{author}{\bibfnamefont{J.~E.} \bibnamefont{Moore}},
  \bibinfo{journal}{Phys. Rev. Lett.} \textbf{\bibinfo{volume}{109}},
  \bibinfo{pages}{017202} (\bibinfo{year}{2012}).

\bibitem[{\citenamefont{Serbyn et~al.}(2013{\natexlab{a}})\citenamefont{Serbyn,
  Papi{\'c}, and Abanin}}]{serbyn2013universal}
\bibinfo{author}{\bibfnamefont{M.}~\bibnamefont{Serbyn}},
  \bibinfo{author}{\bibfnamefont{Z.}~\bibnamefont{Papi{\'c}}},
  \bibnamefont{and} \bibinfo{author}{\bibfnamefont{D.~A.}
  \bibnamefont{Abanin}}, \bibinfo{journal}{Physical review letters}
  \textbf{\bibinfo{volume}{110}}, \bibinfo{pages}{260601}
  (\bibinfo{year}{2013}{\natexlab{a}}).

\bibitem[{\citenamefont{{Swingle}}(2013)}]{Swingle:2013aa}
\bibinfo{author}{\bibfnamefont{B.}~\bibnamefont{{Swingle}}},
  \bibinfo{journal}{ArXiv e-prints}  (\bibinfo{year}{2013}),
  \eprint{1307.0507}.

\bibitem[{\citenamefont{Serbyn et~al.}(2014{\natexlab{a}})\citenamefont{Serbyn,
  Papi\ifmmode~\acute{c}\else \'{c}\fi{}, and Abanin}}]{Serbyn:2014aa}
\bibinfo{author}{\bibfnamefont{M.}~\bibnamefont{Serbyn}},
  \bibinfo{author}{\bibfnamefont{Z.}~\bibnamefont{Papi\ifmmode~\acute{c}\else
  \'{c}\fi{}}}, \bibnamefont{and} \bibinfo{author}{\bibfnamefont{D.~A.}
  \bibnamefont{Abanin}}, \bibinfo{journal}{Phys. Rev. B}
  \textbf{\bibinfo{volume}{90}}, \bibinfo{pages}{174302}
  (\bibinfo{year}{2014}{\natexlab{a}}).

\bibitem[{\citenamefont{Iyer et~al.}(2013)\citenamefont{Iyer, Oganesyan,
  Refael, and Huse}}]{iyer2013many}
\bibinfo{author}{\bibfnamefont{S.}~\bibnamefont{Iyer}},
  \bibinfo{author}{\bibfnamefont{V.}~\bibnamefont{Oganesyan}},
  \bibinfo{author}{\bibfnamefont{G.}~\bibnamefont{Refael}}, \bibnamefont{and}
  \bibinfo{author}{\bibfnamefont{D.~A.} \bibnamefont{Huse}},
  \bibinfo{journal}{Phys. Rev. B 87, 134202}  (\bibinfo{year}{2013}).

\bibitem[{\citenamefont{Kj{\"a}ll et~al.}(2014)\citenamefont{Kj{\"a}ll,
  Bardarson, and Pollmann}}]{kjall2014many}
\bibinfo{author}{\bibfnamefont{J.~A.} \bibnamefont{Kj{\"a}ll}},
  \bibinfo{author}{\bibfnamefont{J.~H.} \bibnamefont{Bardarson}},
  \bibnamefont{and} \bibinfo{author}{\bibfnamefont{F.}~\bibnamefont{Pollmann}},
  \bibinfo{journal}{Phys. Rev. Lett.} \textbf{\bibinfo{volume}{113}},
  \bibinfo{pages}{107204} (\bibinfo{year}{2014}).

\bibitem[{\citenamefont{Laumann et~al.}(2014)\citenamefont{Laumann, Pal, and
  Scardicchio}}]{laumann2014many}
\bibinfo{author}{\bibfnamefont{C.}~\bibnamefont{Laumann}},
  \bibinfo{author}{\bibfnamefont{A.}~\bibnamefont{Pal}}, \bibnamefont{and}
  \bibinfo{author}{\bibfnamefont{A.}~\bibnamefont{Scardicchio}},
  \bibinfo{journal}{Physical review letters} \textbf{\bibinfo{volume}{113}},
  \bibinfo{pages}{200405} (\bibinfo{year}{2014}).

\bibitem[{\citenamefont{Li et~al.}(2015)\citenamefont{Li, Ganeshan, Pixley, and
  Das~Sarma}}]{Li:2015aa}
\bibinfo{author}{\bibfnamefont{X.}~\bibnamefont{Li}},
  \bibinfo{author}{\bibfnamefont{S.}~\bibnamefont{Ganeshan}},
  \bibinfo{author}{\bibfnamefont{J.~H.} \bibnamefont{Pixley}},
  \bibnamefont{and}
  \bibinfo{author}{\bibfnamefont{S.}~\bibnamefont{Das~Sarma}},
  \bibinfo{journal}{Phys. Rev. Lett.} \textbf{\bibinfo{volume}{115}},
  \bibinfo{pages}{186601} (\bibinfo{year}{2015}).

\bibitem[{\citenamefont{Luitz et~al.}(2015)\citenamefont{Luitz, Laflorencie,
  and Alet}}]{luitz2015many}
\bibinfo{author}{\bibfnamefont{D.~J.} \bibnamefont{Luitz}},
  \bibinfo{author}{\bibfnamefont{N.}~\bibnamefont{Laflorencie}},
  \bibnamefont{and} \bibinfo{author}{\bibfnamefont{F.}~\bibnamefont{Alet}},
  \bibinfo{journal}{Phys. Rev. B} \textbf{\bibinfo{volume}{91}},
  \bibinfo{pages}{081103} (\bibinfo{year}{2015}).

\bibitem[{\citenamefont{Tang et~al.}(2015)\citenamefont{Tang, Iyer, and
  Rigol}}]{Tang:2015th}
\bibinfo{author}{\bibfnamefont{B.}~\bibnamefont{Tang}},
  \bibinfo{author}{\bibfnamefont{D.}~\bibnamefont{Iyer}}, \bibnamefont{and}
  \bibinfo{author}{\bibfnamefont{M.}~\bibnamefont{Rigol}},
  \bibinfo{journal}{Phys. Rev. B} \textbf{\bibinfo{volume}{91}},
  \bibinfo{pages}{161109} (\bibinfo{year}{2015}).

\bibitem[{\citenamefont{Singh et~al.}(2016)\citenamefont{Singh, Bardarson, and
  Pollmann}}]{Singh:2016aa}
\bibinfo{author}{\bibfnamefont{R.}~\bibnamefont{Singh}},
  \bibinfo{author}{\bibfnamefont{J.~H.} \bibnamefont{Bardarson}},
  \bibnamefont{and} \bibinfo{author}{\bibfnamefont{F.}~\bibnamefont{Pollmann}},
  \bibinfo{journal}{New Journal of Physics} \textbf{\bibinfo{volume}{18}},
  \bibinfo{pages}{023046} (\bibinfo{year}{2016}).

\bibitem[{\citenamefont{Schreiber et~al.}(2015)\citenamefont{Schreiber,
  Hodgman, Bordia, L{\"u}schen, Fischer, Vosk, Altman, Schneider, and
  Bloch}}]{Schreiber:2015aa}
\bibinfo{author}{\bibfnamefont{M.}~\bibnamefont{Schreiber}},
  \bibinfo{author}{\bibfnamefont{S.~S.} \bibnamefont{Hodgman}},
  \bibinfo{author}{\bibfnamefont{P.}~\bibnamefont{Bordia}},
  \bibinfo{author}{\bibfnamefont{H.~P.} \bibnamefont{L{\"u}schen}},
  \bibinfo{author}{\bibfnamefont{M.~H.} \bibnamefont{Fischer}},
  \bibinfo{author}{\bibfnamefont{R.}~\bibnamefont{Vosk}},
  \bibinfo{author}{\bibfnamefont{E.}~\bibnamefont{Altman}},
  \bibinfo{author}{\bibfnamefont{U.}~\bibnamefont{Schneider}},
  \bibnamefont{and} \bibinfo{author}{\bibfnamefont{I.}~\bibnamefont{Bloch}},
  \bibinfo{journal}{Science} \textbf{\bibinfo{volume}{349}},
  \bibinfo{pages}{842} (\bibinfo{year}{2015}).

\bibitem[{\citenamefont{Kondov et~al.}(2015)\citenamefont{Kondov, McGehee, Xu,
  and DeMarco}}]{Kondov:2015aa}
\bibinfo{author}{\bibfnamefont{S.~S.} \bibnamefont{Kondov}},
  \bibinfo{author}{\bibfnamefont{W.~R.} \bibnamefont{McGehee}},
  \bibinfo{author}{\bibfnamefont{W.}~\bibnamefont{Xu}}, \bibnamefont{and}
  \bibinfo{author}{\bibfnamefont{B.}~\bibnamefont{DeMarco}},
  \bibinfo{journal}{Phys. Rev. Lett.} \textbf{\bibinfo{volume}{114}},
  \bibinfo{pages}{083002} (\bibinfo{year}{2015}).

\bibitem[{\citenamefont{Bordia et~al.}(2016)\citenamefont{Bordia, L\"uschen,
  Hodgman, Schreiber, Bloch, and Schneider}}]{Bordia:2016aa}
\bibinfo{author}{\bibfnamefont{P.}~\bibnamefont{Bordia}},
  \bibinfo{author}{\bibfnamefont{H.~P.} \bibnamefont{L\"uschen}},
  \bibinfo{author}{\bibfnamefont{S.~S.} \bibnamefont{Hodgman}},
  \bibinfo{author}{\bibfnamefont{M.}~\bibnamefont{Schreiber}},
  \bibinfo{author}{\bibfnamefont{I.}~\bibnamefont{Bloch}}, \bibnamefont{and}
  \bibinfo{author}{\bibfnamefont{U.}~\bibnamefont{Schneider}},
  \bibinfo{journal}{Phys. Rev. Lett.} \textbf{\bibinfo{volume}{116}},
  \bibinfo{pages}{140401} (\bibinfo{year}{2016}).

\bibitem[{\citenamefont{{Choi} et~al.}(2016)\citenamefont{{Choi}, {Hild},
  {Zeiher}, {Schau{\ss}}, {Rubio-Abadal}, {Yefsah}, {Khemani}, {Huse}, {Bloch},
  and {Gross}}}]{Choi:2016aa}
\bibinfo{author}{\bibfnamefont{J.-y.} \bibnamefont{{Choi}}},
  \bibinfo{author}{\bibfnamefont{S.}~\bibnamefont{{Hild}}},
  \bibinfo{author}{\bibfnamefont{J.}~\bibnamefont{{Zeiher}}},
  \bibinfo{author}{\bibfnamefont{P.}~\bibnamefont{{Schau{\ss}}}},
  \bibinfo{author}{\bibfnamefont{A.}~\bibnamefont{{Rubio-Abadal}}},
  \bibinfo{author}{\bibfnamefont{T.}~\bibnamefont{{Yefsah}}},
  \bibinfo{author}{\bibfnamefont{V.}~\bibnamefont{{Khemani}}},
  \bibinfo{author}{\bibfnamefont{D.~A.} \bibnamefont{{Huse}}},
  \bibinfo{author}{\bibfnamefont{I.}~\bibnamefont{{Bloch}}}, \bibnamefont{and}
  \bibinfo{author}{\bibfnamefont{C.}~\bibnamefont{{Gross}}},
  \bibinfo{journal}{ArXiv e-prints}  (\bibinfo{year}{2016}),
  \eprint{1604.04178}.

\bibitem[{\citenamefont{{Smith} et~al.}(2015)\citenamefont{{Smith}, {Lee},
  {Richerme}, {Neyenhuis}, {Hess}, {Hauke}, {Heyl}, {Huse}, and
  {Monroe}}}]{Smith:2015aa}
\bibinfo{author}{\bibfnamefont{J.}~\bibnamefont{{Smith}}},
  \bibinfo{author}{\bibfnamefont{A.}~\bibnamefont{{Lee}}},
  \bibinfo{author}{\bibfnamefont{P.}~\bibnamefont{{Richerme}}},
  \bibinfo{author}{\bibfnamefont{B.}~\bibnamefont{{Neyenhuis}}},
  \bibinfo{author}{\bibfnamefont{P.~W.} \bibnamefont{{Hess}}},
  \bibinfo{author}{\bibfnamefont{P.}~\bibnamefont{{Hauke}}},
  \bibinfo{author}{\bibfnamefont{M.}~\bibnamefont{{Heyl}}},
  \bibinfo{author}{\bibfnamefont{D.~A.} \bibnamefont{{Huse}}},
  \bibnamefont{and} \bibinfo{author}{\bibfnamefont{C.}~\bibnamefont{{Monroe}}},
  \bibinfo{journal}{ArXiv e-prints}  (\bibinfo{year}{2015}),
  \eprint{1508.07026}.

\bibitem[{\citenamefont{Serbyn et~al.}(2014{\natexlab{b}})\citenamefont{Serbyn,
  Knap, Gopalakrishnan, Papi\ifmmode~\acute{c}\else \'{c}\fi{}, {Yao}, Laumann,
  Abanin, Lukin, and Demler}}]{Serbyn:2014ek}
\bibinfo{author}{\bibfnamefont{M.}~\bibnamefont{Serbyn}},
  \bibinfo{author}{\bibfnamefont{M.}~\bibnamefont{Knap}},
  \bibinfo{author}{\bibfnamefont{S.}~\bibnamefont{Gopalakrishnan}},
  \bibinfo{author}{\bibfnamefont{Z.}~\bibnamefont{Papi\ifmmode~\acute{c}\else
  \'{c}\fi{}}}, \bibinfo{author}{\bibfnamefont{N.~Y.} \bibnamefont{{Yao}}},
  \bibinfo{author}{\bibfnamefont{C.}~\bibnamefont{Laumann}},
  \bibinfo{author}{\bibfnamefont{D.}~\bibnamefont{Abanin}},
  \bibinfo{author}{\bibfnamefont{M.}~\bibnamefont{Lukin}}, \bibnamefont{and}
  \bibinfo{author}{\bibfnamefont{E.~A.} \bibnamefont{Demler}},
  \bibinfo{journal}{Phys. Rev. Lett.} \textbf{\bibinfo{volume}{113}},
  \bibinfo{pages}{147204} (\bibinfo{year}{2014}{\natexlab{b}}).

\bibitem[{\citenamefont{{Yao} et~al.}(2015)\citenamefont{{Yao}, {Laumann}, and
  {Vishwanath}}}]{Yao:2015aa}
\bibinfo{author}{\bibfnamefont{N.~Y.} \bibnamefont{{Yao}}},
  \bibinfo{author}{\bibfnamefont{C.~R.} \bibnamefont{{Laumann}}},
  \bibnamefont{and}
  \bibinfo{author}{\bibfnamefont{A.}~\bibnamefont{{Vishwanath}}},
  \bibinfo{journal}{ArXiv e-prints}  (\bibinfo{year}{2015}),
  \eprint{1508.06995}.

\bibitem[{\citenamefont{Bahri et~al.}(2015)\citenamefont{Bahri, Vosk, Altman,
  and Vishwanath}}]{Bahri:2015aa}
\bibinfo{author}{\bibfnamefont{Y.}~\bibnamefont{Bahri}},
  \bibinfo{author}{\bibfnamefont{R.}~\bibnamefont{Vosk}},
  \bibinfo{author}{\bibfnamefont{E.}~\bibnamefont{Altman}}, \bibnamefont{and}
  \bibinfo{author}{\bibfnamefont{A.}~\bibnamefont{Vishwanath}},
  \bibinfo{journal}{Nat Commun} \textbf{\bibinfo{volume}{6}}
  (\bibinfo{year}{2015}).

\bibitem[{\citenamefont{Potter et~al.}(2015)\citenamefont{Potter, Vasseur, and
  Parameswaran}}]{Potter:2015ab}
\bibinfo{author}{\bibfnamefont{A.~C.} \bibnamefont{Potter}},
  \bibinfo{author}{\bibfnamefont{R.}~\bibnamefont{Vasseur}}, \bibnamefont{and}
  \bibinfo{author}{\bibfnamefont{S.~A.} \bibnamefont{Parameswaran}},
  \bibinfo{journal}{Phys. Rev. X} \textbf{\bibinfo{volume}{5}},
  \bibinfo{pages}{031033} (\bibinfo{year}{2015}).

\bibitem[{\citenamefont{Chandran and Laumann}(2015)}]{Chandran:2015ac}
\bibinfo{author}{\bibfnamefont{A.}~\bibnamefont{Chandran}} \bibnamefont{and}
  \bibinfo{author}{\bibfnamefont{C.~R.} \bibnamefont{Laumann}},
  \bibinfo{journal}{Phys. Rev. B} \textbf{\bibinfo{volume}{92}},
  \bibinfo{pages}{024301} (\bibinfo{year}{2015}).

\bibitem[{\citenamefont{Vosk et~al.}(2015)\citenamefont{Vosk, Huse, and
  Altman}}]{Vosk:2015aa}
\bibinfo{author}{\bibfnamefont{R.}~\bibnamefont{Vosk}},
  \bibinfo{author}{\bibfnamefont{D.~A.} \bibnamefont{Huse}}, \bibnamefont{and}
  \bibinfo{author}{\bibfnamefont{E.}~\bibnamefont{Altman}},
  \bibinfo{journal}{Phys. Rev. X} \textbf{\bibinfo{volume}{5}},
  \bibinfo{pages}{031032} (\bibinfo{year}{2015}).

\bibitem[{\citenamefont{Huse et~al.}(2013)\citenamefont{Huse, Nandkishore,
  Oganesyan, Pal, and Sondhi}}]{Huse:2013aa}
\bibinfo{author}{\bibfnamefont{D.~A.} \bibnamefont{Huse}},
  \bibinfo{author}{\bibfnamefont{R.}~\bibnamefont{Nandkishore}},
  \bibinfo{author}{\bibfnamefont{V.}~\bibnamefont{Oganesyan}},
  \bibinfo{author}{\bibfnamefont{A.}~\bibnamefont{Pal}}, \bibnamefont{and}
  \bibinfo{author}{\bibfnamefont{S.~L.} \bibnamefont{Sondhi}},
  \bibinfo{journal}{Phys. Rev. B} \textbf{\bibinfo{volume}{88}},
  \bibinfo{pages}{014206} (\bibinfo{year}{2013}).

\bibitem[{\citenamefont{Chandran et~al.}(2014)\citenamefont{Chandran, Khemani,
  Laumann, and Sondhi}}]{Chandran:2014aa}
\bibinfo{author}{\bibfnamefont{A.}~\bibnamefont{Chandran}},
  \bibinfo{author}{\bibfnamefont{V.}~\bibnamefont{Khemani}},
  \bibinfo{author}{\bibfnamefont{C.~R.} \bibnamefont{Laumann}},
  \bibnamefont{and} \bibinfo{author}{\bibfnamefont{S.~L.}
  \bibnamefont{Sondhi}}, \bibinfo{journal}{Phys. Rev. B}
  \textbf{\bibinfo{volume}{89}}, \bibinfo{pages}{144201}
  (\bibinfo{year}{2014}).

\bibitem[{\citenamefont{{Potter} and {Vishwanath}}(2015)}]{Potter:2015aa}
\bibinfo{author}{\bibfnamefont{A.~C.} \bibnamefont{{Potter}}} \bibnamefont{and}
  \bibinfo{author}{\bibfnamefont{A.}~\bibnamefont{{Vishwanath}}},
  \bibinfo{journal}{ArXiv e-prints}  (\bibinfo{year}{2015}),
  \eprint{1506.00592}.

\bibitem[{\citenamefont{{von Keyserlingk} and
  {Sondhi}}(2016)}]{von-Keyserlingk:2016aa}
\bibinfo{author}{\bibfnamefont{C.~W.} \bibnamefont{{von Keyserlingk}}}
  \bibnamefont{and} \bibinfo{author}{\bibfnamefont{S.~L.}
  \bibnamefont{{Sondhi}}}, \bibinfo{journal}{ArXiv e-prints}
  (\bibinfo{year}{2016}), \eprint{1602.02157}.

\bibitem[{\citenamefont{{Else} and {Nayak}}(2016)}]{Else:2016aa}
\bibinfo{author}{\bibfnamefont{D.~V.} \bibnamefont{{Else}}} \bibnamefont{and}
  \bibinfo{author}{\bibfnamefont{C.}~\bibnamefont{{Nayak}}},
  \bibinfo{journal}{ArXiv e-prints}  (\bibinfo{year}{2016}),
  \eprint{1602.04804}.

\bibitem[{\citenamefont{Reif}(1965)}]{Reif:1965aa}
\bibinfo{author}{\bibfnamefont{F.}~\bibnamefont{Reif}},
  \emph{\bibinfo{title}{Fundamentals of statistical and thermal physics}}
  (\bibinfo{publisher}{McGraw-Hill}, \bibinfo{year}{1965}).

\bibitem[{\citenamefont{Deutsch}(1991)}]{deutsch1991quantum}
\bibinfo{author}{\bibfnamefont{J.~M.} \bibnamefont{Deutsch}},
  \bibinfo{journal}{Phys. Rev. A} \textbf{\bibinfo{volume}{43}},
  \bibinfo{pages}{2046} (\bibinfo{year}{1991}).

\bibitem[{\citenamefont{Srednicki}(1994)}]{srednicki1994chaos}
\bibinfo{author}{\bibfnamefont{M.}~\bibnamefont{Srednicki}},
  \bibinfo{journal}{Phys. Rev. E} \textbf{\bibinfo{volume}{50}},
  \bibinfo{pages}{888} (\bibinfo{year}{1994}).

\bibitem[{\citenamefont{Tasaki}(1998)}]{Tasaki:1998aa}
\bibinfo{author}{\bibfnamefont{H.}~\bibnamefont{Tasaki}},
  \bibinfo{journal}{Phys. Rev. Lett.} \textbf{\bibinfo{volume}{80}},
  \bibinfo{pages}{1373} (\bibinfo{year}{1998}).

\bibitem[{\citenamefont{Rigol et~al.}(2008)\citenamefont{Rigol, Dunjko, and
  Olshanii}}]{Rigol:2008bh}
\bibinfo{author}{\bibfnamefont{M.}~\bibnamefont{Rigol}},
  \bibinfo{author}{\bibfnamefont{V.}~\bibnamefont{Dunjko}}, \bibnamefont{and}
  \bibinfo{author}{\bibfnamefont{M.}~\bibnamefont{Olshanii}},
  \bibinfo{journal}{Nature} \textbf{\bibinfo{volume}{452}},
  \bibinfo{pages}{854} (\bibinfo{year}{2008}).

\bibitem[{\citenamefont{{Khlebnikov} and
  {Kruczenski}}(2013)}]{Khlebnikov:2013aa}
\bibinfo{author}{\bibfnamefont{S.}~\bibnamefont{{Khlebnikov}}}
  \bibnamefont{and}
  \bibinfo{author}{\bibfnamefont{M.}~\bibnamefont{{Kruczenski}}},
  \bibinfo{journal}{ArXiv e-prints}  (\bibinfo{year}{2013}),
  \eprint{1312.4612}.

\bibitem[{\citenamefont{Beugeling et~al.}(2014)\citenamefont{Beugeling,
  Moessner, and Haque}}]{Beugeling:2014aa}
\bibinfo{author}{\bibfnamefont{W.}~\bibnamefont{Beugeling}},
  \bibinfo{author}{\bibfnamefont{R.}~\bibnamefont{Moessner}}, \bibnamefont{and}
  \bibinfo{author}{\bibfnamefont{M.}~\bibnamefont{Haque}},
  \bibinfo{journal}{Phys. Rev. E} \textbf{\bibinfo{volume}{89}},
  \bibinfo{pages}{042112} (\bibinfo{year}{2014}).

\bibitem[{\citenamefont{{Sorg} et~al.}(2014)\citenamefont{{Sorg}, {Vidmar},
  {Pollet}, and {Heidrich-Meisner}}}]{Sorg:2014aa}
\bibinfo{author}{\bibfnamefont{S.}~\bibnamefont{{Sorg}}},
  \bibinfo{author}{\bibfnamefont{L.}~\bibnamefont{{Vidmar}}},
  \bibinfo{author}{\bibfnamefont{L.}~\bibnamefont{{Pollet}}}, \bibnamefont{and}
  \bibinfo{author}{\bibfnamefont{F.}~\bibnamefont{{Heidrich-Meisner}}},
  \bibinfo{journal}{\pra} \textbf{\bibinfo{volume}{90}}, \bibinfo{eid}{033606}
  (\bibinfo{year}{2014}), \eprint{1405.5404}.

\bibitem[{\citenamefont{Steinigeweg et~al.}(2014)\citenamefont{Steinigeweg,
  Khodja, Niemeyer, Gogolin, and Gemmer}}]{Steinigeweg:2014aa}
\bibinfo{author}{\bibfnamefont{R.}~\bibnamefont{Steinigeweg}},
  \bibinfo{author}{\bibfnamefont{A.}~\bibnamefont{Khodja}},
  \bibinfo{author}{\bibfnamefont{H.}~\bibnamefont{Niemeyer}},
  \bibinfo{author}{\bibfnamefont{C.}~\bibnamefont{Gogolin}}, \bibnamefont{and}
  \bibinfo{author}{\bibfnamefont{J.}~\bibnamefont{Gemmer}},
  \bibinfo{journal}{Phys. Rev. Lett.} \textbf{\bibinfo{volume}{112}},
  \bibinfo{pages}{130403} (\bibinfo{year}{2014}).

\bibitem[{\citenamefont{Mondaini et~al.}(2016)\citenamefont{Mondaini, Fratus,
  Srednicki, and Rigol}}]{Mondaini:2016aa}
\bibinfo{author}{\bibfnamefont{R.}~\bibnamefont{Mondaini}},
  \bibinfo{author}{\bibfnamefont{K.~R.} \bibnamefont{Fratus}},
  \bibinfo{author}{\bibfnamefont{M.}~\bibnamefont{Srednicki}},
  \bibnamefont{and} \bibinfo{author}{\bibfnamefont{M.}~\bibnamefont{Rigol}},
  \bibinfo{journal}{Phys. Rev. E} \textbf{\bibinfo{volume}{93}},
  \bibinfo{pages}{032104} (\bibinfo{year}{2016}).

\bibitem[{\citenamefont{Huse and Oganesyan}(2013)}]{huse2013phenomenology}
\bibinfo{author}{\bibfnamefont{D.~A.} \bibnamefont{Huse}} \bibnamefont{and}
  \bibinfo{author}{\bibfnamefont{V.}~\bibnamefont{Oganesyan}},
  \bibinfo{journal}{arXiv:1305.4915}  (\bibinfo{year}{2013}).

\bibitem[{\citenamefont{Serbyn et~al.}(2013{\natexlab{b}})\citenamefont{Serbyn,
  Papi{\'c}, and Abanin}}]{serbyn2013local}
\bibinfo{author}{\bibfnamefont{M.}~\bibnamefont{Serbyn}},
  \bibinfo{author}{\bibfnamefont{Z.}~\bibnamefont{Papi{\'c}}},
  \bibnamefont{and} \bibinfo{author}{\bibfnamefont{D.~A.}
  \bibnamefont{Abanin}}, \bibinfo{journal}{Physical review letters}
  \textbf{\bibinfo{volume}{111}}, \bibinfo{pages}{127201}
  (\bibinfo{year}{2013}{\natexlab{b}}).

\bibitem[{\citenamefont{Imbrie}(2014)}]{imbrie2014many}
\bibinfo{author}{\bibfnamefont{J.~Z.} \bibnamefont{Imbrie}},
  \bibinfo{journal}{arXiv preprint arXiv:1403.7837}  (\bibinfo{year}{2014}).

\bibitem[{\citenamefont{Chandran
  et~al.}(2015{\natexlab{a}})\citenamefont{Chandran, Kim, Vidal, and
  Abanin}}]{chandran2015constructing}
\bibinfo{author}{\bibfnamefont{A.}~\bibnamefont{Chandran}},
  \bibinfo{author}{\bibfnamefont{I.~H.} \bibnamefont{Kim}},
  \bibinfo{author}{\bibfnamefont{G.}~\bibnamefont{Vidal}}, \bibnamefont{and}
  \bibinfo{author}{\bibfnamefont{D.~A.} \bibnamefont{Abanin}},
  \bibinfo{journal}{Physical Review B} \textbf{\bibinfo{volume}{91}},
  \bibinfo{pages}{085425} (\bibinfo{year}{2015}{\natexlab{a}}).

\bibitem[{\citenamefont{Ros et~al.}(2015)\citenamefont{Ros, Mueller, and
  Scardicchio}}]{ros2015integrals}
\bibinfo{author}{\bibfnamefont{V.}~\bibnamefont{Ros}},
  \bibinfo{author}{\bibfnamefont{M.}~\bibnamefont{Mueller}}, \bibnamefont{and}
  \bibinfo{author}{\bibfnamefont{A.}~\bibnamefont{Scardicchio}},
  \bibinfo{journal}{Nuclear Physics B} \textbf{\bibinfo{volume}{891}},
  \bibinfo{pages}{420} (\bibinfo{year}{2015}).

\bibitem[{\citenamefont{Monthus}(2016)}]{Monthus:2016aa}
\bibinfo{author}{\bibfnamefont{C.}~\bibnamefont{Monthus}},
  \bibinfo{journal}{Journal of Statistical Mechanics: Theory and Experiment}
  \textbf{\bibinfo{volume}{2016}}, \bibinfo{pages}{033101}
  (\bibinfo{year}{2016}).

\bibitem[{\citenamefont{Rademaker and Ortu\~no}(2016)}]{Rademaker:2016aa}
\bibinfo{author}{\bibfnamefont{L.}~\bibnamefont{Rademaker}} \bibnamefont{and}
  \bibinfo{author}{\bibfnamefont{M.}~\bibnamefont{Ortu\~no}},
  \bibinfo{journal}{Phys. Rev. Lett.} \textbf{\bibinfo{volume}{116}},
  \bibinfo{pages}{010404} (\bibinfo{year}{2016}).

\bibitem[{\citenamefont{Bauer and Nayak}(2013)}]{Bauer:2013jw}
\bibinfo{author}{\bibfnamefont{B.}~\bibnamefont{Bauer}} \bibnamefont{and}
  \bibinfo{author}{\bibfnamefont{C.}~\bibnamefont{Nayak}},
  \bibinfo{journal}{Journal Of Statistical Mechanics-Theory And Experiment}
  \textbf{\bibinfo{volume}{2013}}, \bibinfo{pages}{P09005}
  (\bibinfo{year}{2013}).

\bibitem[{\citenamefont{{Pekker} and {Clark}}(2014)}]{Pekker:2014ab}
\bibinfo{author}{\bibfnamefont{D.}~\bibnamefont{{Pekker}}} \bibnamefont{and}
  \bibinfo{author}{\bibfnamefont{B.~K.} \bibnamefont{{Clark}}},
  \bibinfo{journal}{ArXiv e-prints}  (\bibinfo{year}{2014}),
  \eprint{1410.2224}.

\bibitem[{\citenamefont{Chandran
  et~al.}(2015{\natexlab{b}})\citenamefont{Chandran, Carrasquilla, Kim, Abanin,
  and Vidal}}]{Chandran:2015aa}
\bibinfo{author}{\bibfnamefont{A.}~\bibnamefont{Chandran}},
  \bibinfo{author}{\bibfnamefont{J.}~\bibnamefont{Carrasquilla}},
  \bibinfo{author}{\bibfnamefont{I.~H.} \bibnamefont{Kim}},
  \bibinfo{author}{\bibfnamefont{D.~A.} \bibnamefont{Abanin}},
  \bibnamefont{and} \bibinfo{author}{\bibfnamefont{G.}~\bibnamefont{Vidal}},
  \bibinfo{journal}{Phys. Rev. B} \textbf{\bibinfo{volume}{92}},
  \bibinfo{pages}{024201} (\bibinfo{year}{2015}{\natexlab{b}}).

\bibitem[{\citenamefont{Friesdorf et~al.}(2015)\citenamefont{Friesdorf, Werner,
  Brown, Scholz, and Eisert}}]{Friesdorf:2015aa}
\bibinfo{author}{\bibfnamefont{M.}~\bibnamefont{Friesdorf}},
  \bibinfo{author}{\bibfnamefont{A.~H.} \bibnamefont{Werner}},
  \bibinfo{author}{\bibfnamefont{W.}~\bibnamefont{Brown}},
  \bibinfo{author}{\bibfnamefont{V.~B.} \bibnamefont{Scholz}},
  \bibnamefont{and} \bibinfo{author}{\bibfnamefont{J.}~\bibnamefont{Eisert}},
  \bibinfo{journal}{Phys. Rev. Lett.} \textbf{\bibinfo{volume}{114}},
  \bibinfo{pages}{170505} (\bibinfo{year}{2015}).

\bibitem[{\citenamefont{{Khemani} et~al.}(2015)\citenamefont{{Khemani},
  {Pollmann}, and {Sondhi}}}]{Khemani:2015aa}
\bibinfo{author}{\bibfnamefont{V.}~\bibnamefont{{Khemani}}},
  \bibinfo{author}{\bibfnamefont{F.}~\bibnamefont{{Pollmann}}},
  \bibnamefont{and} \bibinfo{author}{\bibfnamefont{S.~L.}
  \bibnamefont{{Sondhi}}}, \bibinfo{journal}{ArXiv e-prints}
  (\bibinfo{year}{2015}), \eprint{1509.00483}.

\bibitem[{\citenamefont{{Yu} et~al.}(2015)\citenamefont{{Yu}, {Pekker}, and
  {Clark}}}]{yu2015finding}
\bibinfo{author}{\bibfnamefont{X.}~\bibnamefont{{Yu}}},
  \bibinfo{author}{\bibfnamefont{D.}~\bibnamefont{{Pekker}}}, \bibnamefont{and}
  \bibinfo{author}{\bibfnamefont{B.~K.} \bibnamefont{{Clark}}},
  \bibinfo{journal}{ArXiv e-prints}  (\bibinfo{year}{2015}),
  \eprint{1509.01244}.

\bibitem[{\citenamefont{Serbyn et~al.}(2015)\citenamefont{Serbyn,
  Papi\ifmmode~\acute{c}\else \'{c}\fi{}, and Abanin}}]{Serbyn:2015aa}
\bibinfo{author}{\bibfnamefont{M.}~\bibnamefont{Serbyn}},
  \bibinfo{author}{\bibfnamefont{Z.}~\bibnamefont{Papi\ifmmode~\acute{c}\else
  \'{c}\fi{}}}, \bibnamefont{and} \bibinfo{author}{\bibfnamefont{D.~A.}
  \bibnamefont{Abanin}}, \bibinfo{journal}{Phys. Rev. X}
  \textbf{\bibinfo{volume}{5}}, \bibinfo{pages}{041047} (\bibinfo{year}{2015}).

\bibitem[{\citenamefont{Baygan et~al.}(2015)\citenamefont{Baygan, Lim, and
  Sheng}}]{Baygan:2015aa}
\bibinfo{author}{\bibfnamefont{E.}~\bibnamefont{Baygan}},
  \bibinfo{author}{\bibfnamefont{S.~P.} \bibnamefont{Lim}}, \bibnamefont{and}
  \bibinfo{author}{\bibfnamefont{D.~N.} \bibnamefont{Sheng}},
  \bibinfo{journal}{Phys. Rev. B} \textbf{\bibinfo{volume}{92}},
  \bibinfo{pages}{195153} (\bibinfo{year}{2015}).

\bibitem[{\citenamefont{Nandkishore et~al.}(2014)\citenamefont{Nandkishore,
  Gopalakrishnan, and Huse}}]{Nandkishore:2014ys}
\bibinfo{author}{\bibfnamefont{R.}~\bibnamefont{Nandkishore}},
  \bibinfo{author}{\bibfnamefont{S.}~\bibnamefont{Gopalakrishnan}},
  \bibnamefont{and} \bibinfo{author}{\bibfnamefont{D.~A.} \bibnamefont{Huse}},
  \bibinfo{journal}{Phys. Rev. B} \textbf{\bibinfo{volume}{90}},
  \bibinfo{pages}{064203} (\bibinfo{year}{2014}).

\bibitem[{\citenamefont{Johri et~al.}(2015)\citenamefont{Johri, Nandkishore,
  and Bhatt}}]{Johri:2015mz}
\bibinfo{author}{\bibfnamefont{S.}~\bibnamefont{Johri}},
  \bibinfo{author}{\bibfnamefont{R.}~\bibnamefont{Nandkishore}},
  \bibnamefont{and} \bibinfo{author}{\bibfnamefont{R.~N.} \bibnamefont{Bhatt}},
  \bibinfo{journal}{Phys. Rev. Lett.} \textbf{\bibinfo{volume}{114}},
  \bibinfo{pages}{117401} (\bibinfo{year}{2015}).

\bibitem[{\citenamefont{Huse et~al.}(2015)\citenamefont{Huse, Nandkishore,
  Pietracaprina, Ros, and Scardicchio}}]{Huse:2015ys}
\bibinfo{author}{\bibfnamefont{D.~A.} \bibnamefont{Huse}},
  \bibinfo{author}{\bibfnamefont{R.}~\bibnamefont{Nandkishore}},
  \bibinfo{author}{\bibfnamefont{F.}~\bibnamefont{Pietracaprina}},
  \bibinfo{author}{\bibfnamefont{V.}~\bibnamefont{Ros}}, \bibnamefont{and}
  \bibinfo{author}{\bibfnamefont{A.}~\bibnamefont{Scardicchio}},
  \bibinfo{journal}{Phys. Rev. B} \textbf{\bibinfo{volume}{92}},
  \bibinfo{pages}{014203} (\bibinfo{year}{2015}).

\bibitem[{\citenamefont{{Hyatt} et~al.}(2016)\citenamefont{{Hyatt}, {Garrison},
  {Potter}, and {Bauer}}}]{Hyatt:2016aa}
\bibinfo{author}{\bibfnamefont{K.}~\bibnamefont{{Hyatt}}},
  \bibinfo{author}{\bibfnamefont{J.~R.} \bibnamefont{{Garrison}}},
  \bibinfo{author}{\bibfnamefont{A.~C.} \bibnamefont{{Potter}}},
  \bibnamefont{and} \bibinfo{author}{\bibfnamefont{B.}~\bibnamefont{{Bauer}}},
  \bibinfo{journal}{ArXiv e-prints}  (\bibinfo{year}{2016}),
  \eprint{1601.07184}.

\bibitem[{\citenamefont{Greif et~al.}(2016)\citenamefont{Greif, Parsons,
  Mazurenko, Chiu, Blatt, Huber, Ji, and Greiner}}]{Greif953}
\bibinfo{author}{\bibfnamefont{D.}~\bibnamefont{Greif}},
  \bibinfo{author}{\bibfnamefont{M.~F.} \bibnamefont{Parsons}},
  \bibinfo{author}{\bibfnamefont{A.}~\bibnamefont{Mazurenko}},
  \bibinfo{author}{\bibfnamefont{C.~S.} \bibnamefont{Chiu}},
  \bibinfo{author}{\bibfnamefont{S.}~\bibnamefont{Blatt}},
  \bibinfo{author}{\bibfnamefont{F.}~\bibnamefont{Huber}},
  \bibinfo{author}{\bibfnamefont{G.}~\bibnamefont{Ji}}, \bibnamefont{and}
  \bibinfo{author}{\bibfnamefont{M.}~\bibnamefont{Greiner}},
  \bibinfo{journal}{Science} \textbf{\bibinfo{volume}{351}},
  \bibinfo{pages}{953} (\bibinfo{year}{2016}).

\bibitem[{\citenamefont{Huse et~al.}(2014)\citenamefont{Huse, Nandkishore, and
  Oganesyan}}]{huse2014phenomenology}
\bibinfo{author}{\bibfnamefont{D.~A.} \bibnamefont{Huse}},
  \bibinfo{author}{\bibfnamefont{R.}~\bibnamefont{Nandkishore}},
  \bibnamefont{and}
  \bibinfo{author}{\bibfnamefont{V.}~\bibnamefont{Oganesyan}},
  \bibinfo{journal}{Physical Review B} \textbf{\bibinfo{volume}{90}},
  \bibinfo{pages}{174202} (\bibinfo{year}{2014}).

\bibitem[{\citenamefont{Chen et~al.}(2015)\citenamefont{Chen, Yu, Cho, Clark,
  and Fradkin}}]{Chen:2015aa}
\bibinfo{author}{\bibfnamefont{X.}~\bibnamefont{Chen}},
  \bibinfo{author}{\bibfnamefont{X.}~\bibnamefont{Yu}},
  \bibinfo{author}{\bibfnamefont{G.~Y.} \bibnamefont{Cho}},
  \bibinfo{author}{\bibfnamefont{B.~K.} \bibnamefont{Clark}}, \bibnamefont{and}
  \bibinfo{author}{\bibfnamefont{E.}~\bibnamefont{Fradkin}},
  \bibinfo{journal}{Phys. Rev. B} \textbf{\bibinfo{volume}{92}},
  \bibinfo{pages}{214204} (\bibinfo{year}{2015}).

\bibitem[{\citenamefont{Devakul and Singh}(2015)}]{Devakul:2015aa}
\bibinfo{author}{\bibfnamefont{T.}~\bibnamefont{Devakul}} \bibnamefont{and}
  \bibinfo{author}{\bibfnamefont{R.~R.~P.} \bibnamefont{Singh}},
  \bibinfo{journal}{Phys. Rev. Lett.} \textbf{\bibinfo{volume}{115}},
  \bibinfo{pages}{187201} (\bibinfo{year}{2015}).

\bibitem[{\citenamefont{Baldwin et~al.}(2016)\citenamefont{Baldwin, Laumann,
  Pal, and Scardicchio}}]{Baldwin:2016aa}
\bibinfo{author}{\bibfnamefont{C.~L.} \bibnamefont{Baldwin}},
  \bibinfo{author}{\bibfnamefont{C.~R.} \bibnamefont{Laumann}},
  \bibinfo{author}{\bibfnamefont{A.}~\bibnamefont{Pal}}, \bibnamefont{and}
  \bibinfo{author}{\bibfnamefont{A.}~\bibnamefont{Scardicchio}},
  \bibinfo{journal}{Phys. Rev. B} \textbf{\bibinfo{volume}{93}},
  \bibinfo{pages}{024202} (\bibinfo{year}{2016}).

\bibitem[{\citenamefont{Serbyn and Moore}(2016)}]{Serbyn:2016aa}
\bibinfo{author}{\bibfnamefont{M.}~\bibnamefont{Serbyn}} \bibnamefont{and}
  \bibinfo{author}{\bibfnamefont{J.~E.} \bibnamefont{Moore}},
  \bibinfo{journal}{Phys. Rev. B} \textbf{\bibinfo{volume}{93}},
  \bibinfo{pages}{041424} (\bibinfo{year}{2016}).

\bibitem[{\citenamefont{Srednicki}(1999)}]{srednicki1999thermal}
\bibinfo{author}{\bibfnamefont{M.}~\bibnamefont{Srednicki}},
  \bibinfo{journal}{Journal of Physics A: Mathematical and General}
  \textbf{\bibinfo{volume}{32}}, \bibinfo{pages}{1163} (\bibinfo{year}{1999}).

\bibitem[{\citenamefont{{D'Alessio} et~al.}(2015)\citenamefont{{D'Alessio},
  {Kafri}, {Polkovnikov}, and {Rigol}}}]{DAlessio:2015aa}
\bibinfo{author}{\bibfnamefont{L.}~\bibnamefont{{D'Alessio}}},
  \bibinfo{author}{\bibfnamefont{Y.}~\bibnamefont{{Kafri}}},
  \bibinfo{author}{\bibfnamefont{A.}~\bibnamefont{{Polkovnikov}}},
  \bibnamefont{and} \bibinfo{author}{\bibfnamefont{M.}~\bibnamefont{{Rigol}}},
  \bibinfo{journal}{ArXiv e-prints}  (\bibinfo{year}{2015}),
  \eprint{1509.06411}.

\bibitem[{\citenamefont{Grover}(2014)}]{Grover:2014aa}
\bibinfo{author}{\bibfnamefont{T.}~\bibnamefont{Grover}}
  (\bibinfo{year}{2014}), \eprint{1405.1471}.

\bibitem[{\citenamefont{{Potter} et~al.}(2016)\citenamefont{{Potter},
  {Morimoto}, and {Vishwanath}}}]{Potter:2016aa}
\bibinfo{author}{\bibfnamefont{A.~C.} \bibnamefont{{Potter}}},
  \bibinfo{author}{\bibfnamefont{T.}~\bibnamefont{{Morimoto}}},
  \bibnamefont{and}
  \bibinfo{author}{\bibfnamefont{A.}~\bibnamefont{{Vishwanath}}},
  \bibinfo{journal}{ArXiv e-prints}  (\bibinfo{year}{2016}),
  \eprint{1602.05194}.

\bibitem[{\citenamefont{De~Roeck et~al.}(2016)\citenamefont{De~Roeck,
  Huveneers, M\"uller, and Schiulaz}}]{De-Roeck:2016aa}
\bibinfo{author}{\bibfnamefont{W.}~\bibnamefont{De~Roeck}},
  \bibinfo{author}{\bibfnamefont{F.}~\bibnamefont{Huveneers}},
  \bibinfo{author}{\bibfnamefont{M.}~\bibnamefont{M\"uller}}, \bibnamefont{and}
  \bibinfo{author}{\bibfnamefont{M.}~\bibnamefont{Schiulaz}},
  \bibinfo{journal}{Phys. Rev. B} \textbf{\bibinfo{volume}{93}},
  \bibinfo{pages}{014203} (\bibinfo{year}{2016}).

\bibitem[{\citenamefont{Mondragon-Shem
  et~al.}(2015)\citenamefont{Mondragon-Shem, Pal, Hughes, and
  Laumann}}]{Mondragon-Shem:2015aa}
\bibinfo{author}{\bibfnamefont{I.}~\bibnamefont{Mondragon-Shem}},
  \bibinfo{author}{\bibfnamefont{A.}~\bibnamefont{Pal}},
  \bibinfo{author}{\bibfnamefont{T.~L.} \bibnamefont{Hughes}},
  \bibnamefont{and} \bibinfo{author}{\bibfnamefont{C.~R.}
  \bibnamefont{Laumann}}, \bibinfo{journal}{Phys. Rev. B}
  \textbf{\bibinfo{volume}{92}}, \bibinfo{pages}{064203}
  (\bibinfo{year}{2015}).

\bibitem[{\citenamefont{De~Luca and Scardicchio}(2013)}]{de2013ergodicity}
\bibinfo{author}{\bibfnamefont{A.}~\bibnamefont{De~Luca}} \bibnamefont{and}
  \bibinfo{author}{\bibfnamefont{A.}~\bibnamefont{Scardicchio}},
  \bibinfo{journal}{Europhys. Lett.} \textbf{\bibinfo{volume}{101}},
  \bibinfo{pages}{37003} (\bibinfo{year}{2013}).

\bibitem[{\citenamefont{Bakr et~al.}(2009)\citenamefont{Bakr, Gillen, Peng,
  Folling, and Greiner}}]{Bakr:2009aa}
\bibinfo{author}{\bibfnamefont{W.~S.} \bibnamefont{Bakr}},
  \bibinfo{author}{\bibfnamefont{J.~I.} \bibnamefont{Gillen}},
  \bibinfo{author}{\bibfnamefont{A.}~\bibnamefont{Peng}},
  \bibinfo{author}{\bibfnamefont{S.}~\bibnamefont{Folling}}, \bibnamefont{and}
  \bibinfo{author}{\bibfnamefont{M.}~\bibnamefont{Greiner}},
  \bibinfo{journal}{Nature} \textbf{\bibinfo{volume}{462}}, \bibinfo{pages}{74}
  (\bibinfo{year}{2009}).

\bibitem[{\citenamefont{Bakr et~al.}(2010)\citenamefont{Bakr, Peng, Tai, Ma,
  Simon, Gillen, F{\"o}lling, Pollet, and Greiner}}]{Bakr547}
\bibinfo{author}{\bibfnamefont{W.~S.} \bibnamefont{Bakr}},
  \bibinfo{author}{\bibfnamefont{A.}~\bibnamefont{Peng}},
  \bibinfo{author}{\bibfnamefont{M.~E.} \bibnamefont{Tai}},
  \bibinfo{author}{\bibfnamefont{R.}~\bibnamefont{Ma}},
  \bibinfo{author}{\bibfnamefont{J.}~\bibnamefont{Simon}},
  \bibinfo{author}{\bibfnamefont{J.~I.} \bibnamefont{Gillen}},
  \bibinfo{author}{\bibfnamefont{S.}~\bibnamefont{F{\"o}lling}},
  \bibinfo{author}{\bibfnamefont{L.}~\bibnamefont{Pollet}}, \bibnamefont{and}
  \bibinfo{author}{\bibfnamefont{M.}~\bibnamefont{Greiner}},
  \bibinfo{journal}{Science} \textbf{\bibinfo{volume}{329}},
  \bibinfo{pages}{547} (\bibinfo{year}{2010}), ISSN \bibinfo{issn}{0036-8075}.

\bibitem[{\citenamefont{Sherson et~al.}(2010)\citenamefont{Sherson, Weitenberg,
  Endres, Cheneau, Bloch, and Kuhr}}]{Sherson:2010aa}
\bibinfo{author}{\bibfnamefont{J.~F.} \bibnamefont{Sherson}},
  \bibinfo{author}{\bibfnamefont{C.}~\bibnamefont{Weitenberg}},
  \bibinfo{author}{\bibfnamefont{M.}~\bibnamefont{Endres}},
  \bibinfo{author}{\bibfnamefont{M.}~\bibnamefont{Cheneau}},
  \bibinfo{author}{\bibfnamefont{I.}~\bibnamefont{Bloch}}, \bibnamefont{and}
  \bibinfo{author}{\bibfnamefont{S.}~\bibnamefont{Kuhr}},
  \bibinfo{journal}{Nature} \textbf{\bibinfo{volume}{467}}, \bibinfo{pages}{68}
  (\bibinfo{year}{2010}).

\bibitem[{\citenamefont{Edwards and Thouless}(1972)}]{Edwards:1972aa}
\bibinfo{author}{\bibfnamefont{J.~T.} \bibnamefont{Edwards}} \bibnamefont{and}
  \bibinfo{author}{\bibfnamefont{D.~J.} \bibnamefont{Thouless}},
  \bibinfo{journal}{Journal of Physics C: Solid State Physics}
  \textbf{\bibinfo{volume}{5}}, \bibinfo{pages}{807} (\bibinfo{year}{1972}).

\bibitem[{\citenamefont{Thouless}(1974)}]{Thouless:1974aa}
\bibinfo{author}{\bibfnamefont{D.}~\bibnamefont{Thouless}},
  \bibinfo{journal}{Physics Reports} \textbf{\bibinfo{volume}{13}},
  \bibinfo{pages}{93 } (\bibinfo{year}{1974}), ISSN \bibinfo{issn}{0370-1573}.

\bibitem[{\citenamefont{Abrahams et~al.}(1979)\citenamefont{Abrahams, Anderson,
  Licciardello, and Ramakrishnan}}]{Abrahams:1979ud}
\bibinfo{author}{\bibfnamefont{E.}~\bibnamefont{Abrahams}},
  \bibinfo{author}{\bibfnamefont{P.~W.} \bibnamefont{Anderson}},
  \bibinfo{author}{\bibfnamefont{D.~C.} \bibnamefont{Licciardello}},
  \bibnamefont{and} \bibinfo{author}{\bibfnamefont{T.~V.}
  \bibnamefont{Ramakrishnan}}, \bibinfo{journal}{Phys. Rev. Lett.}
  \textbf{\bibinfo{volume}{42}}, \bibinfo{pages}{673} (\bibinfo{year}{1979}).

\bibitem[{\citenamefont{Abrahams}(2010)}]{Abrahams2010}
\bibinfo{author}{\bibfnamefont{E.}~\bibnamefont{Abrahams}},
  \emph{\bibinfo{title}{50 years of Anderson localization}},
  vol.~\bibinfo{volume}{24} (\bibinfo{publisher}{World Scientific},
  \bibinfo{year}{2010}).

\bibitem[{\citenamefont{Bera et~al.}(2015)\citenamefont{Bera, Schomerus,
  Heidrich-Meisner, and Bardarson}}]{Bera:2015aa}
\bibinfo{author}{\bibfnamefont{S.}~\bibnamefont{Bera}},
  \bibinfo{author}{\bibfnamefont{H.}~\bibnamefont{Schomerus}},
  \bibinfo{author}{\bibfnamefont{F.}~\bibnamefont{Heidrich-Meisner}},
  \bibnamefont{and} \bibinfo{author}{\bibfnamefont{J.~H.}
  \bibnamefont{Bardarson}}, \bibinfo{journal}{Phys. Rev. Lett.}
  \textbf{\bibinfo{volume}{115}}, \bibinfo{pages}{046603}
  (\bibinfo{year}{2015}).

\bibitem[{\citenamefont{Bar~Lev et~al.}(2015)\citenamefont{Bar~Lev, Cohen, and
  Reichman}}]{Bar-Lev:2015aa}
\bibinfo{author}{\bibfnamefont{Y.}~\bibnamefont{Bar~Lev}},
  \bibinfo{author}{\bibfnamefont{G.}~\bibnamefont{Cohen}}, \bibnamefont{and}
  \bibinfo{author}{\bibfnamefont{D.~R.} \bibnamefont{Reichman}},
  \bibinfo{journal}{Phys. Rev. Lett.} \textbf{\bibinfo{volume}{114}},
  \bibinfo{pages}{100601} (\bibinfo{year}{2015}).

\bibitem[{\citenamefont{Agarwal et~al.}(2015)\citenamefont{Agarwal,
  Gopalakrishnan, Knap, M\"uller, and Demler}}]{Agarwal:2015aa}
\bibinfo{author}{\bibfnamefont{K.}~\bibnamefont{Agarwal}},
  \bibinfo{author}{\bibfnamefont{S.}~\bibnamefont{Gopalakrishnan}},
  \bibinfo{author}{\bibfnamefont{M.}~\bibnamefont{Knap}},
  \bibinfo{author}{\bibfnamefont{M.}~\bibnamefont{M\"uller}}, \bibnamefont{and}
  \bibinfo{author}{\bibfnamefont{E.}~\bibnamefont{Demler}},
  \bibinfo{journal}{Phys. Rev. Lett.} \textbf{\bibinfo{volume}{114}},
  \bibinfo{pages}{160401} (\bibinfo{year}{2015}).

\bibitem[{\citenamefont{Gopalakrishnan
  et~al.}(2016)\citenamefont{Gopalakrishnan, Agarwal, Demler, Huse, and
  Knap}}]{Gopalakrishnan:2016aa}
\bibinfo{author}{\bibfnamefont{S.}~\bibnamefont{Gopalakrishnan}},
  \bibinfo{author}{\bibfnamefont{K.}~\bibnamefont{Agarwal}},
  \bibinfo{author}{\bibfnamefont{E.~A.} \bibnamefont{Demler}},
  \bibinfo{author}{\bibfnamefont{D.~A.} \bibnamefont{Huse}}, \bibnamefont{and}
  \bibinfo{author}{\bibfnamefont{M.}~\bibnamefont{Knap}},
  \bibinfo{journal}{Phys. Rev. B} \textbf{\bibinfo{volume}{93}},
  \bibinfo{pages}{134206} (\bibinfo{year}{2016}).

\bibitem[{\citenamefont{{Kerala Varma} et~al.}(2015)\citenamefont{{Kerala
  Varma}, {Lerose}, {Pietracaprina}, {Goold}, and
  {Scardicchio}}}]{Kerala-Varma:2015aa}
\bibinfo{author}{\bibfnamefont{V.}~\bibnamefont{{Kerala Varma}}},
  \bibinfo{author}{\bibfnamefont{A.}~\bibnamefont{{Lerose}}},
  \bibinfo{author}{\bibfnamefont{F.}~\bibnamefont{{Pietracaprina}}},
  \bibinfo{author}{\bibfnamefont{J.}~\bibnamefont{{Goold}}}, \bibnamefont{and}
  \bibinfo{author}{\bibfnamefont{A.}~\bibnamefont{{Scardicchio}}},
  \bibinfo{journal}{ArXiv e-prints}  (\bibinfo{year}{2015}),
  \eprint{1511.09144}.

\bibitem[{\citenamefont{{Khemani} and Huse}()}]{Khemani:aa}
\bibinfo{author}{\bibfnamefont{V.}~\bibnamefont{{Khemani}}} \bibnamefont{and}
  \bibinfo{author}{\bibfnamefont{D.~A.} \bibnamefont{Huse}},
  \bibinfo{note}{private communication}.

\end{thebibliography}

\appendix

\section{Spectral representation from the ETH ansatz}
\label{App:ETHDerivation}
Consider the correlation function:
\begin{align}
\label{Eq:IDef}
	\mathcal{I} = \bra{E_\alpha}  [H,\lstar]^2 \ket{E_\alpha}
\end{align}
in the eigenstate $\ket{E_\alpha}$. In this appendix, we use the ETH ansatz to show that:
\begin{align}
	\mathcal{I} = \int_{-\infty}^\infty d\omega\, e^{-\beta \omega/2} \omega^2 |f(E_\alpha,\omega)|^2
\end{align}
where $f$ is the spectral function associated with $\lstar$.

Inserting a complete set of states in Eq.~\eqref{Eq:IDef} and using the ETH ansatz for $\lstar$:
\begin{align}
\label{Eq:CommSumForm}
 \mathcal{I} = \sum_{\beta} \frac{\omega^2}{\rho(\bar{E})} r^2_{\alpha \beta} |f(\bar{E}, \omega)|^2
\end{align}
where $\bar{E} = (E_\alpha + E_\beta)/2$ and $\omega = E_\alpha - E_\beta$. 
Assuming that $f$ is a smooth function on the scale of the many-body level spacing $\Delta E \sim1/\rho(\bar{E})$, the sum can be replaced by the integral:
\begin{equation}
\sum_{\beta} r_{\alpha \beta}^2 \to \int dE_\beta\rho(E_\beta)
\end{equation}
Note that $\rho(E) \sim e^{S(E)}$, where $S(E)$ is the microcanonical entropy at energy $E$, and the many body density of states has units of inverse energy. On further changing the integration variable to $\omega$, we obtain:
\begin{align}
\label{Eq:IStepMid}
 \mathcal{I} = \int_{-\infty}^{\infty} d\omega  \frac{\rho(E_\alpha - \omega)}{\rho(E_\alpha - \omega/2)} \omega^2 |f(E_\alpha - \omega/2, \omega)|^2.
\end{align}

The $f$ function is expected to decay exponentially at large $\omega$ on a scale $\omega_0$ that is at most $O(L^0)$.
As $E_\alpha \sim O(L^d)$, $\omega_0 \ll E_\alpha$ and the domain of integration in Eq.~\eqref{Eq:IStepMid} may be restricted to $\omega \ll E_\alpha$ to obtain a very good approximation to $\mathcal{I}$.
Quantitatively, the error in this approximation is exponentially small in $L^d/\omega_0$.

Expanding $S(E)$ and $f$ for $\omega \ll E_\alpha$:
\begin{align*}
S(E_\alpha - \omega) - S(E_\alpha - \omega/2) &\approx -\omega \beta/2 + O(1/L^d) \\
f(E_\alpha - \omega/2, \omega) &\approx f(E_\alpha , \omega) + O(1/L^d)
\end{align*}
where $\beta$ is the inverse temperature $\left.\frac{dS}{dE}\right|_{E_\alpha} = \beta$.
Putting it all together, we obtain the desired result:
\begin{align}
 \mathcal{I} = \int_{-\infty}^{\infty} d\omega\, e^{-\beta \omega/2} \omega^2 |f(E_\alpha , \omega)|^2.
\end{align}
In the same way, we can show from 
\begin{align}
\bra{E} \lstar \lstar \ket{E} - \bra{E} \lstar \ket{E}^2 \sim O(1),
\end{align}
we get:
\begin{align}
\label{Eq:ProbDensity}
\int d\omega e^{-\beta \omega/2} |f(E,\omega)|^2 \sim O(1).
\end{align}
\end{document}